\documentclass[aps,pre,reprint,onecolumn,groupedaddress]{revtex4-2}

\usepackage{graphicx}
\usepackage{amsmath}
\usepackage{color,soul}
\usepackage{caption}
\usepackage{subcaption}
\usepackage{float}
\usepackage{multirow}

\usepackage{amssymb}
\usepackage{pifont}

\begin{document}


\title{Unrestricted component count in multiphase lattice Boltzmann: a fugacity-based approach}

\author{Muzammil Soomro}
\email{msoomro@psu.edu}
\author{Luis F. Ayala}

\affiliation{Department of Energy and Mineral Engineering, The Pennsylvania State University, University Park, PA 16802, USA}

\date{\today}

\begin{abstract}
Studies of multiphase fluids utilizing the lattice Boltzmann method (LBM) are typically severely restricted by the number of components or chemical species being modeled. This restriction is particularly pronounced for multiphase systems exhibiting partial miscibility and significant interfacial mass exchange, which is a common occurrence in realistic multiphase systems. Modeling such systems becomes increasingly complex as the number of chemical species increases due to the increased role of molecular interactions and the types of thermodynamic behavior that become possible. The recently introduced fugacity-based LBM [M. Soomro, L. F. Ayala, C. Peng, and O. M. Ayala, Phys. Rev. E 107, 015304 (2023)] has provided a thermodynamically-consistent modeling platform for multicomponent, partially-miscible LBM simulations. However, until now, this fugacity-based LB model had lacked a comprehensive demonstration of its ability to accurately reproduce thermodynamic behavior beyond binary mixtures and to remove any restrictions in a number of components for multiphase LBM. In this paper, we closely explore these fugacity-based LBM capabilities by showcasing comprehensive, thermodynamically-consistent simulations of multiphase mixtures of up to ten chemical components. The paper begins by validating the model against the Young-Laplace equation for a droplet composed of three components. The model is then applied to study mixtures with a range of component numbers from one to six, showing agreement with rigorous thermodynamic predictions and demonstrating linear scaling of computational time with the number of components. We further investigate--which has been previously absent in LB literature--ternary systems in detail, by exploring a wide range of temperature, pressure, and overall composition conditions to produce various characteristic ternary diagrams. In addition, the model is shown to be unrestricted in the number of phases, as demonstrated through simulations of a three-component three-phase equilibrium case. The paper concludes by demonstrating simulations of a ten-component, realistic hydrocarbon mixture, achieving excellent agreement with thermodynamics for both flat interface vapor-liquid equilibrium and curved interface spinodal decomposition cases. This study represents a significant expansion of the scope and capabilities of multiphase LBM simulations that encompass multiphase systems of keen interest in engineering.
\end{abstract}

\maketitle

\section{Introduction}
\label{secIntro}

The lattice Boltzmann method (LBM) has proven to be a powerful tool to simulate flow, particularly for multiphase systems. One of the features that make the LBM such an attractive tool is its ability to handle multiple phases with different chemical components. However, with very few exceptions, multicomponent applications of multiphase LBM have mostly been restricted to (1) binary-only (mostly) or ternary-only systems, and (2) immiscible phases. This severely limits the scope of multiphase LBM as complex fluids with multiple components, beyond just binary and ternary systems, are ubiquitous in industrial and natural systems and interfacial mass transfer, or partial-miscibility is an important consideration in these systems. In our recently published paper \cite{Soomro2023}, we develop an LBM formulation based on the fugacity thermodynamic property, which is capable to overcome the chemical component restriction in LBM and enables thermodynamically-consistent, partially-miscible simulations. However, in that study, we only provided a comprehensive analysis of the thermodynamic behavior of binary systems. In the current paper, we provide proof that our fugacity-based formulation, along with some suggested modifications, can be used to accurately simulate partially-miscible fluids with any number of components, under a wide range of conditions, and in precise agreement with thermodynamic predictions. We also present a comprehensive analysis of the thermodynamic behavior expected of partially-miscible ternary systems. 

The flow of complex fluids consisting of multiple components is of great importance in various applications, including geological carbon dioxide sequestration, solute extraction in liquid-liquid systems, and hydrocarbon recovery. Hydrocarbons, in particular, are an important example of such fluids since they can be composed of hundreds of distinct chemical components \cite{pedersen2015}. However, modeling these systems poses significant challenges as they are known to exhibit complex phase behavior. Depending on temperature, pressure, and composition conditions, they can exist as a single phase or multiple phases, where the degree of miscibility of these phases is a crucial consideration. Although immiscible flow models are commonly embraced because they significantly simplify the resulting hydrodynamic and thermodynamic equations, they only represent an idealized subset of all the partial-miscibility scenarios that exist in practice. In reality, phases exhibit at least some degree of interfacial mass transfer and are partially miscible. Given the immense importance of these complex fluids, there is a need for partially-miscible flow models that can handle a large number of components. This is challenging due to the increase in complexity of phase behavior with the addition of every single component.

Any attempt to model such systems using the LBM needs to couple multiphase LBM with a robust thermodynamic model. One popular approach to incorporating multicomponent thermodynamics is to use equations of state (EOSs) since they enable the estimation of fugacity \cite{sandler2006}, a well-established thermodynamic property that is a proxy to chemical potential and can thus be considered as the potential driving the flux of chemical species. As a result, fugacity can be deployed in lieu of chemical potential to arrive at thermodynamic equilibrium conditions via the iso-fugacity criterion. Modern cubic EOSs are examples of EOSs that have been refined over the years to accurately replicate multicomponent and multiphase thermodynamics. The van der Waals (vdW) EOS was the first cubic EOS, but it is known to be quantitatively inaccurate for modeling phase behavior. More modern cubic EOSs, such as the Soave-Redlich-Kwong (SRK) EOS \cite{Soave1972} and Peng-Robinson (PR) EOS \cite{Peng1976}, are better known for their accuracy. The PR EOS is particularly prevalent in modeling hydrocarbon systems. Additionally, a cubic EOS can be extended to multi-component systems by utilizing vdW random mixing rules \cite{Kwak1986}.

The LBM can be extended to multiphase systems via a number of approaches, such as the pseudo-potential model \cite{Shan1993,Shan1994}, and free energy model \cite{Swift1995,Swift1996}.
The pseudo-potential model introduces a force at the mesoscopic scale which replicates intermolecular interactions and through this force, phase separation can be triggered \cite{Shan1993,Shan1994}. This model allows for the use of a variety of cubic EOSs for single component systems \cite{Yuan2006}, although it is a well-known problem that pseudo-potential LBM remains unable to be fully consistent with thermodynamics i.e., at equilibrium, pseudo-potential LBM cannot reach a state where chemical potentials are uniform within the system \cite{Li2012}. This problem can be mitigated, but not fully removed, by introducing tuning parameters that can be adjusted to approach the iso-chemical potential criterion \cite{Kupershtokh2009}. Alternatively, the free energy LBM introduces multiphase phenomena at the macroscopic level through a pressure tensor based on a functional of the Helmholtz free energy \cite{Luo1998}. Although this model takes into account macroscopic thermodynamics, early free energy formulations also violated the iso-chemical potential criterion of equilibrium. This violation is attributed to discretization errors but can be resolved through the implementation of a ``well-balanced" free energy formulation \cite{Guo2021}. Free energy models also enable the use of cubic EOSs \cite{Mazloomi2015PRLSC,Mazloomi2015PRESC,Siebert2014SC,Wen2020SC,Zhang2022WB}. However, it had largely been applied to single-component systems and had not been fully generalized to multicomponent systems \cite{Soomro2023}. Both pseudo-potential and free energy models have been extended beyond single-component systems, with much of the earlier work focused on immiscible rather than partially-miscible systems. And even for these immiscible models, the emphasis had been on binary \cite{Martys1996,Wang2022,Liu20162C,Scarbolo20132C} or ternary mixtures \cite{Liang20163C,Lamura19993C,Wohrwag2018,He2020}, and generalizing them to mixtures of four or more components has proven to be challenging \cite{Yuan2022MC}. There are models that can be generalized for any number of components \cite{Yuan2022MC,Zheng2020MC,Semprebon2016MC}, but in the absence of realistic multicomponent, multiphase thermodynamics and thus cannot be extended to partially-miscible systems. He et al. \cite{He2020b} introduced a ternary LBM model that could handle both immiscible and fully miscible fluids, and was able to simulate cases like the coalescence of two miscible droplets and the rise of gas bubbles in water to reach a water-air interface. Nonetheless, all fluid phases in this model must be either immiscible or fully miscible and never partially-miscible.

Some multicomponent studies have attempted to capture partially-miscible systems. One widely-used model, developed by Bao and Schaefer \cite{Bao2013}, extends the pseudo-potential model to multiple components and is shown to capture phase compositions in an air-water system. However, in their approach, they apply an EOS to each component to obtain independent pressure estimations for each component. This approach can lead to significant thermodynamic inconsistencies because the pressure of a real fluid is a property of the macroscopic phase and not a component. When multiple components are present, mixing rules should be accounted for and a single EOS used to obtain the pressure of a phase. Gong et al. \cite{Gong2014} proposed a pseudo-potential model that splits the mesoscopic force between components. However, their force-splitting strategy was designed empirically and cannot be derived from or supported by physical considerations \cite{Peng2021}. Peng et al. \cite{Peng2021} proposed another force-splitting approach for the pseudo-potential model and designed their force split based on the equality of fugacities between phases. They accurately simulated binary vapor-liquid systems, but their approach remains limited to binary systems and requires a tuning parameter to fully achieve equality of fugacities. Ridl and Wagner \cite{Ridl2018} introduced a multicomponent free energy model for vdW fluids. They were able to perform a comprehensive analysis of binary phase behavior, including two-component two-phase and two-component three-phase simulations. However, their model remains restricted to the vdW EOS and did not provide any direct path for generalization to other modern EOSs. They also show that some of their multiphase simulations require additional corrections when unable to achieve uniform chemical potential throughout the domain. In our recent work (Soomro et al. \cite{Soomro2023}), we proposed a free energy LBM model for partially-miscible systems based on component fugacity, which can be used with any multicomponent EOS, such as the SRK and PR EOSs (vdW included). By incorporating the well-balanced LBM formulation \cite{Guo2021}, Soomro et al.'s model fulfills the iso-chemical potential criterion without any tuning or additional correction. Using this fugacity-based formulation, a comprehensive analysis of binary phase behavior is demonstrated and a sample three-component system was simulated. 

Currently, there is a notable lack of partially-miscible, or even immiscible, LBM studies that utilize a large number of components. From the LBM literature cited in this section, the largest number of components utilized is five, which pertains only to immiscible systems \cite{Zheng2020MC}. For partially-miscible systems, the largest number of components is three, with a lack of a comprehensive analysis \cite{Ridl2018,Gong2014,Bao2013}.  In this paper, we aim to fill this gap by utilizing the fugacity-based LBM \cite{Soomro2023} and introducing an approximation for the interfacial tension parameters when multiple components are present. With this approach, we analyze the phase behavior of partially-miscible systems with a large number of components, including realistic mixtures of up to ten components and a detailed exploration of the rich phase behavior displayed by ternary systems. The remainder of this paper is structured as follows. Section \ref{secMethod} provides a description of the methodology employed in this study. Specifically, we present an overview of the fugacity-based LBM, an approximation to the interfacial tension parameter, and an overview of the PR EOS and its associated fugacity expression. A quantitative method to establish the values of the interfacial tension parameter for each component present in the system is also presented. In Section \ref{secResults}, we show the results obtained using the fugacity-based LBM. First, we demonstrate the agreement of our model with the Young-Laplace equation in Section \ref{secLaplace}. Next, we simulate vapor-liquid equilibrium for mixtures containing one to six components in Section \ref{secMulticomp}. We then conduct a more comprehensive analysis of ternary systems in Section \ref{secTernaryVLE}, demonstrating the ternary phase behavior across a range of temperature, pressure, and overall compositions. In addition, we showcase a case of three-component three-phase equilibrium in Section \ref{secThreePhase}. Finally, in Section \ref{sec10Comp} we examine the vapor-liquid equilibrium and spinodal decomposition simulations of a ten-component hydrocarbon mixture. The main conclusions of the paper are then discussed in Section \ref{secConclusion}.

\section{Methodology}
\label{secMethod}

\subsection{The fugacity-based LBM}

In this study, we deploy the fugacity-based LBM \cite{Soomro2023}. The lattice Boltzmann equation (LBE) utilized in this study differs from the standard LBE in two key aspects: (1) it is designed for multiple components and (2) it employs the well-balanced formulation, which features a different expression for the equilibrium distribution function and the forcing term. The resulting multicomponent, well-balanced LBE is given by Equation \ref{eqLBEMC}:

\begin{subequations}
    \label{eqLBEMC}
    \begin{equation}
    \label{eqLBEMCa}
            g_{\alpha,i}\left(\mathbf{r}+\mathbf{e}_{\alpha}\delta t,t+\delta t\right)-g_{\alpha,i}\left(\mathbf{r},t\right)=-\frac{1}{\tau}\left[g_{\alpha,i}\left(\mathbf{r},t\right)-g_{\alpha,i}^{(eq)}\left(\mathbf{r},t\right)\right]+\left[1-\frac{1}{2\tau}\right]F_{\alpha,i}\left(\mathbf{r},t\right)\delta t,
    \end{equation}
    \begin{equation}
    \label{eqLBEMCb}
          g_{\alpha,i}^{(eq)}=  \begin{cases}
    \rho_i-\left(1-w_0\right)\rho_{i}^c+w_0\rho_i \left[\frac{\mathbf{u}\cdot \mathbf{e}_{\alpha}}{c_s^2}+\frac{\left(\mathbf{u}\cdot \mathbf{e}_{\alpha}\right)^2}{2c_s^4}-\frac{\mathbf{u}\cdot \mathbf{u}}{2c_s^2}\right] & \text{if $\alpha=0$ }\\
    w_\alpha\rho_{i}^c+w_\alpha\rho_i\left[\frac{\mathbf{u}\cdot \mathbf{e}_{\alpha}}{c_s^2}+\frac{\left(\mathbf{u}\cdot \mathbf{e}_{\alpha}\right)^2}{2c_s^4}-\frac{\mathbf{u}\cdot \mathbf{u}}{2c_s^2}\right] & \text{if $\alpha\ne0$}\\
    \end{cases},
    \end{equation}
    \begin{equation}
    \label{eqLBEMCc}
            F_{\alpha,i}=\mathbf{F}_i\cdot w_{\alpha}\left[\frac{\mathbf{e}_{\alpha}-\mathbf{u}}{c_s^2}+\frac{\left(\mathbf{u}\cdot \mathbf{e}_{\alpha}\right)\mathbf{e}_{\alpha}}{c_s^4}\right]+\nabla\rho_i\cdot w_{\alpha} \left[-\mathbf{u}+ \frac{\left(\mathbf{u}\cdot \mathbf{e}_{\alpha}\right)\mathbf{e}_{\alpha}}{c_s^2} +\frac{1}{2}\left(\frac{\mathbf{e}_\alpha^2}{c_s^2}-D\right)\mathbf{u} \right],
    \end{equation}
\end{subequations}

where the subscript `$i$' refers to the $i$-th component and the subscript `$\alpha$' refers to the direction along the discrete lattice velocity. $g_{\alpha,i}$, $g_{\alpha,i}^{(eq)}$, and $F_{\alpha,i}$ are the distribution function, equilibrium distribution function, and forcing term, respectively, for direction $\alpha$ and component $i$. $\mathbf{e}_\alpha$ and $w_\alpha$ are the lattice velocity and weighing parameter, respectively, for direction $\alpha$. $\rho_i$ and $\mathbf{F}_i$ are the density and body force, respectively, for component $i$ (note that $\rho_i=m_i/V$, where $m_i$ is the mass of component $i$ in an element and $V$ is the volume of that element). $\mathbf{r}$, $\mathbf{u}$, $t$, $\tau$, $c_s$, and $D$ are the position vector, macroscopic velocity vector, time, relaxation time, speed of sound, and spacial dimension of the problem, respectively. $\rho_{i}^c$ is a numerical constant for component $i$, which is set to 0 \cite{Guo2021}.

The body force, $\mathbf{F}_i$, in Equation \ref{eqLBEMCc} can be written in terms of the chemical potential of component $i$, denoted by $\mu_i$. This is commonly referred to as the ``potential form" of the body force. The chemical potential can be decomposed into contributions from the bulk fluid, denoted by subscript `$B$', and the interface, denoted by subscript `$I$'. The potential form of the body force is given by Equation \ref{eqForceMCPot}:

\begin{equation}
    \label{eqForceMCPot}
    \mathbf{F}_i=-\tilde{\rho}_i\nabla\mu_{B,i}-\tilde{\rho}_i\nabla\mu_{I,i},
\end{equation}

where $\tilde{\rho}_i$ is the molar density of component $i$, which can be computed as $\tilde{\rho}_i=n_i/V$, where $n_i$ is the amount, in moles, of component $i$ in an element and $V$ is the volume of that element. The gradient of the bulk chemical potential for component `i' can be obtained through the gradient of the fugacity of component `i', which is a readily available property for any EOS \cite{Soomro2023,Lewis1901}. The interface chemical potential can be obtained through the free energy of inhomogeneous systems \cite{Carey1980,Ridl2018}.  By making these substitutions, we can obtain the final expression for the component force, which is given by Equation \ref{eqForce2},

\begin{equation}
\label{eqForce2}
    \mathbf{F}_i=-\tilde{\rho}_i RT\nabla\ln{f_i}+\tilde{\rho}_i\sum_{j=1}^{N}\nabla\left(\kappa_{ij}\ \nabla^2\tilde{\rho}_j\right).
\end{equation}

Here $f_i$ is the fugacity of component $i$, $R$ is the universal gas constant, $T$ is the temperature, and $N$ is the number of components in the mixture. $\kappa_{ij}$ is a parameter that arises from the molecular interactions between component $i$ and $j$ and controls the interfacial tension. $\kappa_{ij}$ represents the same energetic interactions as $a_{ij}$, the attraction term for a component pair $i$-$j$ in a cubic EOS. Thus, the same mixing rules used for $a_{ij}$ can be applied to $\kappa_{ij}$ \cite{Carey1980}:

\begin{equation}
    \label{eqkappa}
    \kappa_{ij}=\sqrt{\kappa_i\kappa_j},
\end{equation}

where $\kappa_i$ is the interfacial tension parameter for a pure component $i$. Using Equation \ref{eqkappa}, the body force can be expressed as:

\begin{equation}
\label{eqforcefinal}
    \mathbf{F}_i=-\tilde{\rho}_i RT\nabla\ln{f_i}+\tilde{\rho}_i\sum_{j=1}^{N}\nabla\left(\sqrt{\kappa_i\kappa_j}\ \nabla^2\tilde{\rho}_j\right).
\end{equation}

For a more detailed explanation of the fugacity-based LBM methodology and its implementation, readers are referred to Ref. \cite{Soomro2023}.

\subsection{Interfacial tension parameter}

In the free energy LBM, the desired interfacial tension in a system is often achieved by adjusting the value of $\kappa_i$. In single component systems, only one parameter needs to be tuned to achieve the desired interfacial tension. However, for multicomponent systems, there are $N$ parameters that need to be tuned. Different combinations of $\kappa_i$ can produce the same value of interfacial tension, which can lead to a potentially over-specified system. A case can be made that since $\kappa_i$ controls the interfacial tension in a pure system, it should be calibrated based on a pure system. For instance, if we intend to simulate a multicomponent system of $N$ components at a specific temperature $T$, the value of $\kappa_i$ for each component should be tuned in a pure component system to reach the actual pure component interfacial tension for that component at $T$. These values of $\kappa_i$ can then be utilized in the multi-component system at $T$. This can provide a unique, albeit cumbersome, approach to specify $\kappa_i$. Unfortunately, this approach can also fail in certain cases where $T$ is greater than the critical temperature of some of the components, as those pure components will exist as a single phase. 

To avoid the issue of over-specification and reduce the parameters to tune, an approximation for $\kappa_i$ can be used. Since, $\kappa_i$ and $a_i$ (attraction parameter for component $i$ in cubic EOSs) are both a function of the same molecular interactions, it is reasonable to assume that $\kappa_i\propto a_i$. This proportionality is already implied in Equation \ref{eqkappa}. By taking the $\kappa_i$ of one of the components in the mixture as a reference, we can obtain $\kappa_i$ for the remaining components as shown in Equation \ref{eqkapparef}:

\begin{equation}
\label{eqkapparef}
    \kappa_i=\kappa_i^{ref} \frac{a_i}{a_{i}^{ref}}.
\end{equation}

Cubic EOSs all have similar expressions for the attraction term:

\begin{equation}
\label{eqattrac}
    a_i=\Omega_a\frac{R^2T_{c,i}^2}{p_{c,i}},
\end{equation}

where $T_{c,i}$ and $p_{c,i}$ are the critical temperature and critical pressure of component $i$, respectively. $\Omega_a$ is a constant that varies between EOSs. Using Equations \ref{eqkapparef} and \ref{eqattrac}, we can arrive with and an expression for $\kappa_i$, given a reference value, as shown in Equation \ref{eqkappafinal},

\begin{equation}
\label{eqkappafinal}
    \kappa_i=\kappa_i^{ref} \left(\frac{T_{c,i}}{T_{c,i}^{ref}}\right)^2  \frac{p_{c,i}^{ref}}{p_{c,i}}.
\end{equation}

\subsection{EOS Selection}
\label{secMethodPR}

In this study, we will be utilizing hydrocarbon mixtures as our test system. The PR EOS is widely recognized for its ability to accurately model hydrocarbon behavior. Thus, we have chosen to employ it for our analyses. However, it should be noted that the fugacity-based LBM is agnostic to EOS selection and can be deployed with any preferred EOS for the system of interest. For an $N$-component mixture, the PR EOS is given by Equation \ref{eqPREOSMC}:

\begin{equation}
    \label{eqPREOSMC}
    p=\frac{RT}{\tilde{v}-b_m}-\frac{(a\alpha)_m}{\tilde{v}^2+2b_m\tilde{v}-b_m^2},
\end{equation}

where $\tilde{v}$ is the molar volume ($\tilde{v}=1/\tilde{\rho}$). The mixing rules for the parameters $a$, $\alpha$, and $b$ are provided by Equations \ref{eqMixinga} and \ref{eqMixingb} \cite{Kwak1986},

\begin{equation}
\label{eqMixinga}
    (a\alpha)_m=\sum_{i=1}^{N}\sum_{j=1}^{N}x_ix_j\sqrt{(a\alpha)_i(a\alpha)_j}\left(1-\delta_{ij} \right),
\end{equation}

\begin{equation}
\label{eqMixingb}
    b_m=\sum_{i=1}^{N}x_ib_i.
\end{equation}

Here, $x_i$ is the mole fraction, or composition, of component $i$, and $\delta_{ij}$ is the binary interaction parameter between component $i$ and $j$. Moreover, the parameters $a_i$, $b_i$, and $\alpha_i$ are defined as follows: $a_i=0.457235529\frac{R^2T_{c,i}^2}{p_{c,i}}$, $ b_i=0.077796074\frac{RT_{c,i}}{p_{c,i}}$, and

\begin{equation*}
    \alpha_i=
    \begin{cases}
    \left[1+\left(0.374640+1.54226\omega_i-0.26992\omega_i^2 \right) \left(1-T_{r,i}^{0.5} \right) \right]^2 & \text{if $\omega_i\le0.49$ }\\
    \left[1+\left(0.379642+1.48503\omega_i-0.164423\omega_i^2+0.016666\omega_i^3 \right) \left(1-T_{r,i}^{0.5} \right) \right]^2 & \text{if $\omega_i>0.49$}\\
    \end{cases},
\end{equation*}

where $\omega_i$ is the acentric factor for component $i$ and $T_{r,i}=\frac{T}{T_{c,i}}$. Using Equation \ref{eqPREOSMC}, the fugacity expression for the PR EOS can be derived from the definition of fugacity. The PR fugacity is shown in Equation \ref{eqfugPR}.

\begin{equation}
    \label{eqfugPR}
    \ln{\left[\frac{f_i}{x_ip}\right]} = \frac{b_i}{b_{m}}\left[\frac{p\tilde{v}}{RT}-1\right]-\ln{\left[\frac{\left(\tilde{v}-b_m\right)p}{RT}\right]}+\frac{(a\alpha)_{m}}{2\sqrt{2}b_mRT}\left[\frac{b_i}{b_m}-\frac{2}{(a\alpha)_m}\sum_{j=1}^{N_c}x_j(a\alpha)_{ij}\right]\ln{\left[\frac{\tilde{v}+\left(1+\sqrt{2}\right)b_m}{\tilde{v}+\left(1-\sqrt{2}\right)b_m}\right]}
\end{equation}

For the derivation of Equation \ref{eqfugPR}, readers can refer to Appendix C of Ref. \cite{Soomro2023}. In this study, we will frequently utilize the term ``overall composition". The overall composition of component $i$ is defined as the ratio of the number of moles of component $i$ in the entire system to the total number of moles of all components in the system. This quantity will be denoted by $z_i$.

\section{Results}
\label{secResults}

In this section, we test our formulation for several different cases. First, we conduct capillary pressure measurements in a 2D suspended droplet case for different droplet radii and $\kappa^{ref}_i$ values to assess their agreement with the Young-Laplace equation. Next, we investigate vapor-liquid equilibrium (VLE) cases for flat interface simulations comprising of one to six components.  Subsequently, we delve deeper into ternary systems, employing data from numerous VLE cases to construct ternary diagrams, and present a three-component three-phase simulation. Finally, we demonstrate the effectiveness of our formulation with a realistic 10-component hydrocarbon mixture, by showcasing flat interface VLE cases and 2D cases of spinodal decomposition. All simulations were carried out in a periodic computational domain, employing a D2Q9 lattice and using the PR EOS. Mixtures will be composed of several components, and component properties and abbreviations are provided in Table \ref{tabCompProp} unless otherwise stated (as will be the case in Section \ref{secThreePhase}). The component C7+ in Table \ref{tabCompProp}, represents a (non-unique) grouping of all components with a chain length of 7 and above, which is to be used in Section \ref{sec10Comp}.

\begin{table}[h]
\caption{The properties of relevant components used in the LBM simulations.}
\label{tabCompProp}
\centering
\begin{tabular}{lcccc}
\hline \hline
Component & Critical Pressure (bar)\ \  & Critical Temperature (K)\ \  & Acentric factor\ \  & Molar Mass (g/mol)\ \  \\
\hline
Carbon   Dioxide (CO2) & 73.843 & 304.39 & 0.2667 & 44.010  \\
Methane (C1)           & 45.947 & 190.74 & 0.0104 & 16.043  \\
Ethane (C2)            & 48.711 & 305.51 & 0.0979 & 30.070  \\
Propane (C3)           & 42.472 & 370.03 & 0.1522 & 44.097  \\
iso-Butane (iC4)       & 36.397 & 408.03 & 0.1852 & 58.123  \\
n-Butane (C4)          & 37.963 & 425.34 & 0.1995 & 58.123  \\
iso-Pentane (iC5)       & 33.812 & 460.61 & 0.2280 & 72.150  \\
n-Pentane (C5)         & 33.688 & 469.89 & 0.2514 & 72.150  \\
n-Hexane (C6)          & 30.123 & 507.56 & 0.2994 & 86.177  \\
C7+                    & 21.043 & 617.78 & 0.4898 & 142.285 
\\
\hline \hline
\end{tabular}
\end{table}

\subsection{Young-Laplace Equation Validation}
\label{secLaplace}
In this case, we aim to test whether our model, with the $\kappa_i\propto a_i$ approximation, results in capillary pressure predictions in compliance with the Young-Laplace equation. We will simulate a 2D static droplet inside a vapor phase. For this, we use a system of C1, C2, and C3 at a temperature of 250 K, initialized at a pressure of 50 bar. The size of the computational domain is $200\times 200$ ($n_x\times n_y$) and the relaxation time $\tau=1.0$. The binary interaction parameters between all component pairs are 0. The relevant conversions between lattice units and physical units are established by fixing the universal gas constant and the attraction parameter, co-volume, and molar mass for C1 to the following values in lattice units: $R=1$, $a_{C1}=2/49$, $b_{C1}=2/21$, and $M_{C1}=1$.
To initialize the density of each component, we use Equation \ref{eqInit2D},

\begin{equation}
    \label{eqInit2D}
    \rho_i(x,y,t=0)=\frac{\rho_{i,L}+\rho_{i,V}}{2}-\frac{\rho_{i,L}-\rho_{i,V}}{2}\tanh{\left\{\frac{2 \left[\sqrt{\left(x-x_c\right)^2+\left(y-y_c\right)^2}-R\right]}{W}\right\}},
\end{equation}

where $W$ is the initial interface width set to 4, $x_c$ and $y_c$ are the $x$ and $y$ coordinates of the center of the droplet set to $x_c=n_x/2$ and $y_c=n_y/2$, and $R$ is the radius of the droplet at initial conditions. $\rho_{i, V}$ and $\rho_{i, L}$ are the densities of component `$i$' in the vapor phase and liquid phase, respectively. These densities are set to the equilibrium densities at $T=250\ K$ and $p=50\ bar$ obtained through a flash calculation (with an overall composition: $z_{C1}=0.5$, $z_{C2}=0.25$, and $z_{C3}=0.25$). We initialize the droplet using different values of $R$ (20, 30, 40, and 50 in lattice units), and repeat the process for three distinct values of $\kappa_i^{ref}$ (0.02, 0.04, and 0.06 in lattice units), where C1 is selected as the reference component. It is worth noting that in addition to $\rho_{i, V}$ and $\rho_{i, L}$, a flash calculation also prescribes a phase saturation that fixes the value of $R$. Therefore, adjusting the value of $R$ in our simulations effectively changes the value of $z_i$ but this has no effect on the equilibrium values of $\rho_{i,L}$ and $\rho_{i,V}$. However, the density profile will deviate from the one prescribed by Equation \ref{eqInit2D} for two reasons, First, we initialize the densities according to a flash calculation, which assumes a flat interface without capillary pressure. However, in the LBM simulations, there is a curved interface. Second, the ``tanh'' profile is only any approximation to the density profile across the interface.

Each simulation is run for 1,000,000 time-steps to achieve equilibrium. The Young-Laplace equation for a 2D droplet, given by Equation \ref{eqLaplace}, relates the capillary pressure, $\Delta p$, to the radius of the droplet, $R$, and the interfacial tension, $\sigma$:

\begin{equation}
    \label{eqLaplace}
    \Delta p = \frac{\sigma}{R}.
\end{equation}

\begin{figure}[h]
    \centering
    \includegraphics[scale=0.65]{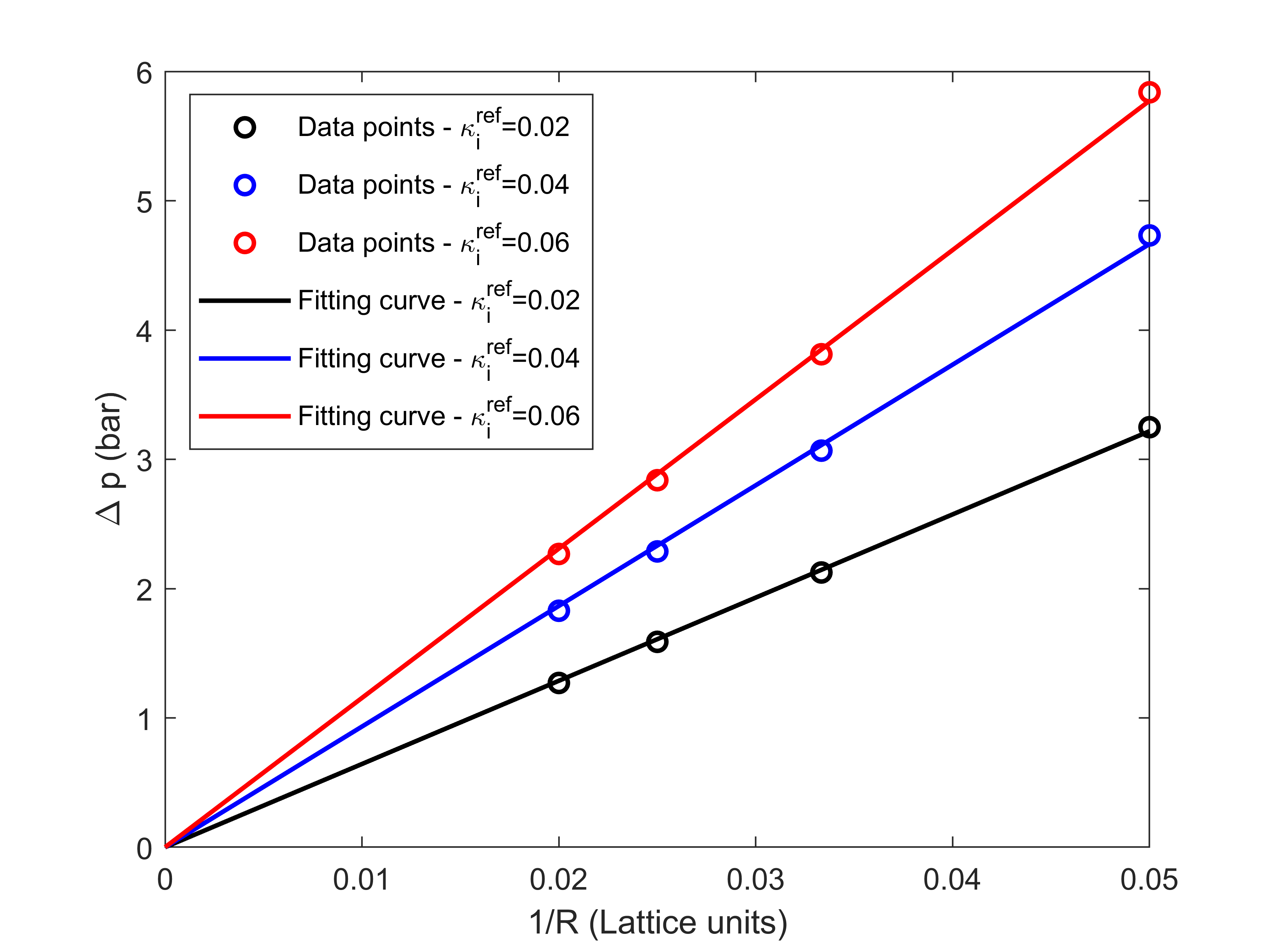}
    \caption{The capillary pressure from each simulation vs the inverse of droplet radius for that simulation. The dots represent the data points obtained from the simulation and the solid lines represent the linear fitting curves.}
    \label{figLaplaceTest}
\end{figure}

As per Equation \ref{eqLaplace}, the capillary pressure is proportional to $1/R$. To test this, we measure the capillary pressure values obtained from each simulation and plot them against their corresponding $1/R$ values. This is repeated for every $\kappa_i^{ref}$ value. The results are presented in Figure \ref{figLaplaceTest}.

It can be seen from Figure \ref{figLaplaceTest}, that the $\Delta p$ values obtained from the LBM simulations are directly proportional to $1/R$, for each value of $\kappa_i^{ref}$. This demonstrates that the simulations are in agreement with the Young-Laplace equation.

\subsection{Multicomponent Vapor-Liquid Equilibrium}
\label{secMulticomp}
To demonstrate the generalizability of our approach, we simulate vapor-liquid equilibrium (VLE) for two phases separated by a flat interface and with up to six components. Six cases are carried out and the components used, the temperature, and initial pressure in these simulations are shown in Table \ref{tabSixComp}. All mixtures used in the six cases are equimolar in terms of the total moles in the system, i.e., all components have the same $z_i$. 

\begin{table}[H]
\caption{Description of the case simulations to be studied.}
\label{tabSixComp}
\centering
\begin{tabular}{lccc}
\hline \hline
Case & Components used\ \  & Temperature (K)\ \  & Pressure (bar)\ \  \\
\hline
1   & C1                     & 177.24 & 30.02  \\
2   & C1, C2                 & 215.00 & 32.00  \\
3   & C1, C2, C3             & 250.00 & 35.00  \\
4   & C1, C2, C3, C4         & 300.00 & 40.00  \\
5   & C1, C2, C3, C4, C5     & 350.00 & 45.00  \\
6   & C1, C2, C3, C4, C5, C6 & 400.00 & 50.00  
\\
\hline \hline
\end{tabular}
\end{table}

For all simulations, the relevant conversions between lattice units and physical units are established by fixing the universal gas constant and the attraction parameter, co-volume, and molar mass for C1 to the following values in lattice units: $R=1$, $a_{C1}=2/49$, $b_{C1}=2/21$, and $M_{C1}=1$. The binary interaction parameters between all component pairs are 0. The relaxation time $\tau=1.0$, and the interfacial tension parameter for the reference component, chosen to be C1, is $\kappa_i^{ref}=0.02$. The size of the computational domain is $400\times 2$ ($n_x\times n_y$), and the density of each component along the $x$ direction is initialized as shown in Equation \ref{eqInit1D} (the domain will be symmetric in the $y$ direction),

\begin{equation}
    \label{eqInit1D}
    \rho_i(x,t=0)=\rho_{i,V}+\frac{\rho_{i,L}-\rho_{i,V}}{2}\left[\tanh{\left(\frac{2\left(x-\frac{S_V}{2}n_x\right)}{W}\right)}-\tanh{\left(\frac{2\left(x-\left(1-\frac{S_V}{2}\right)n_x\right)}{W}\right)}\right].
\end{equation}

\begin{figure}[H]
    \centering
        \begin{subfigure}{0.49\textwidth}
            \centering
            \caption{}
            \includegraphics[width=\textwidth]{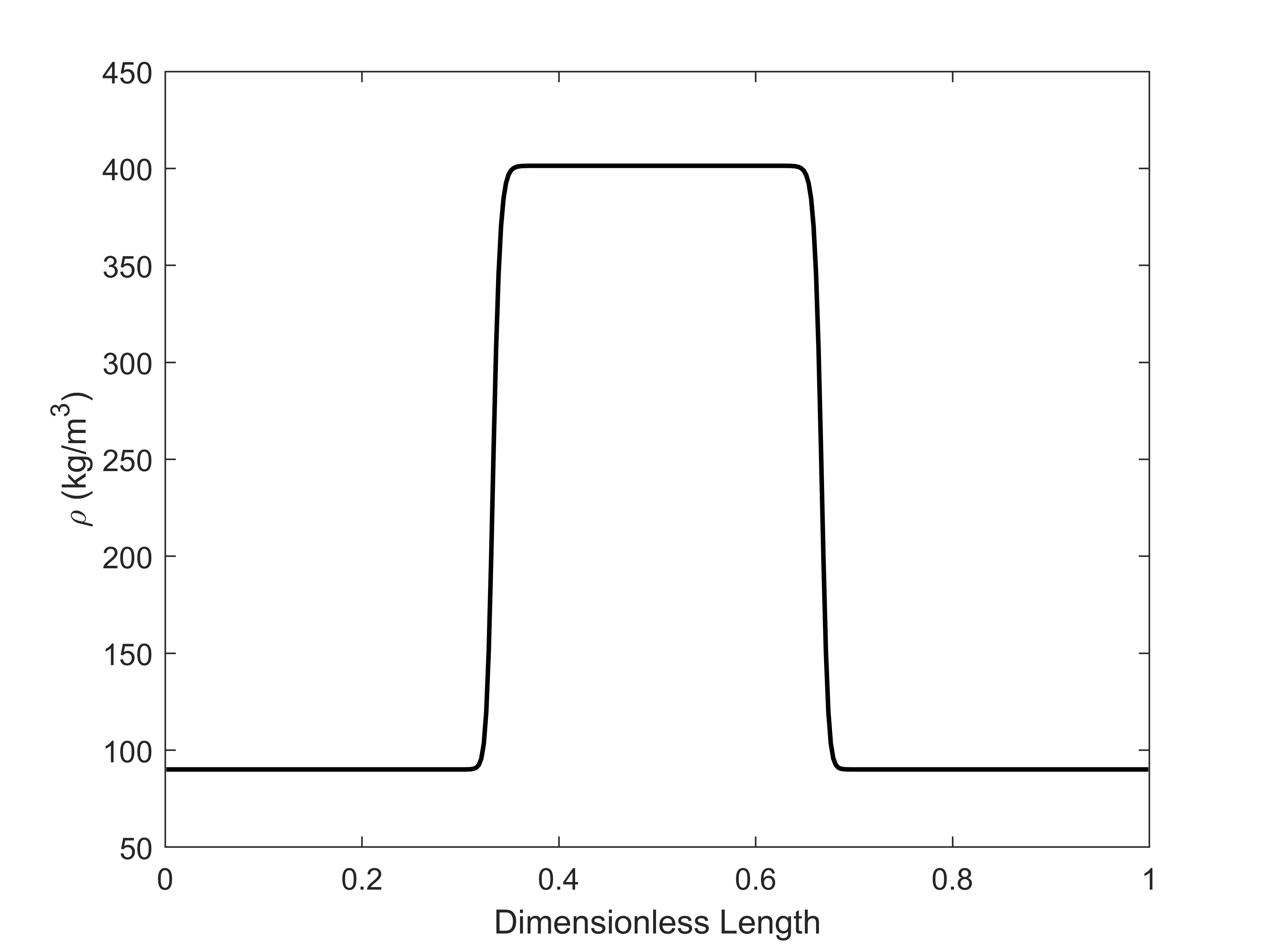}
        \end{subfigure}%
        \hfill
        \begin{subfigure}{0.49\textwidth}
            \centering
            \caption{}
            \includegraphics[width=\textwidth]{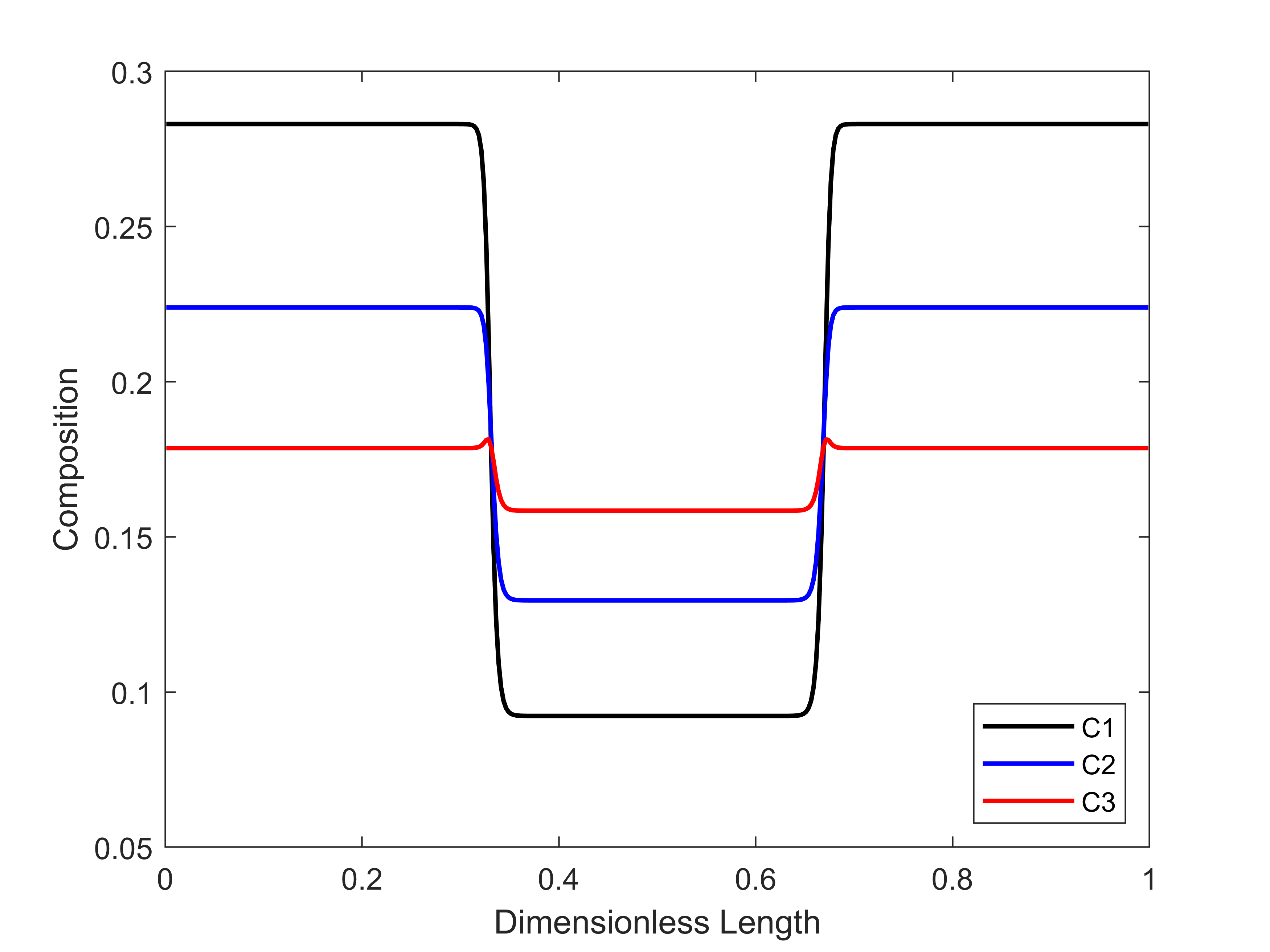}
        \end{subfigure}
        \hfill        
        \begin{subfigure}{0.49\textwidth}
            \centering
            \caption{}
            \includegraphics[width=\textwidth]{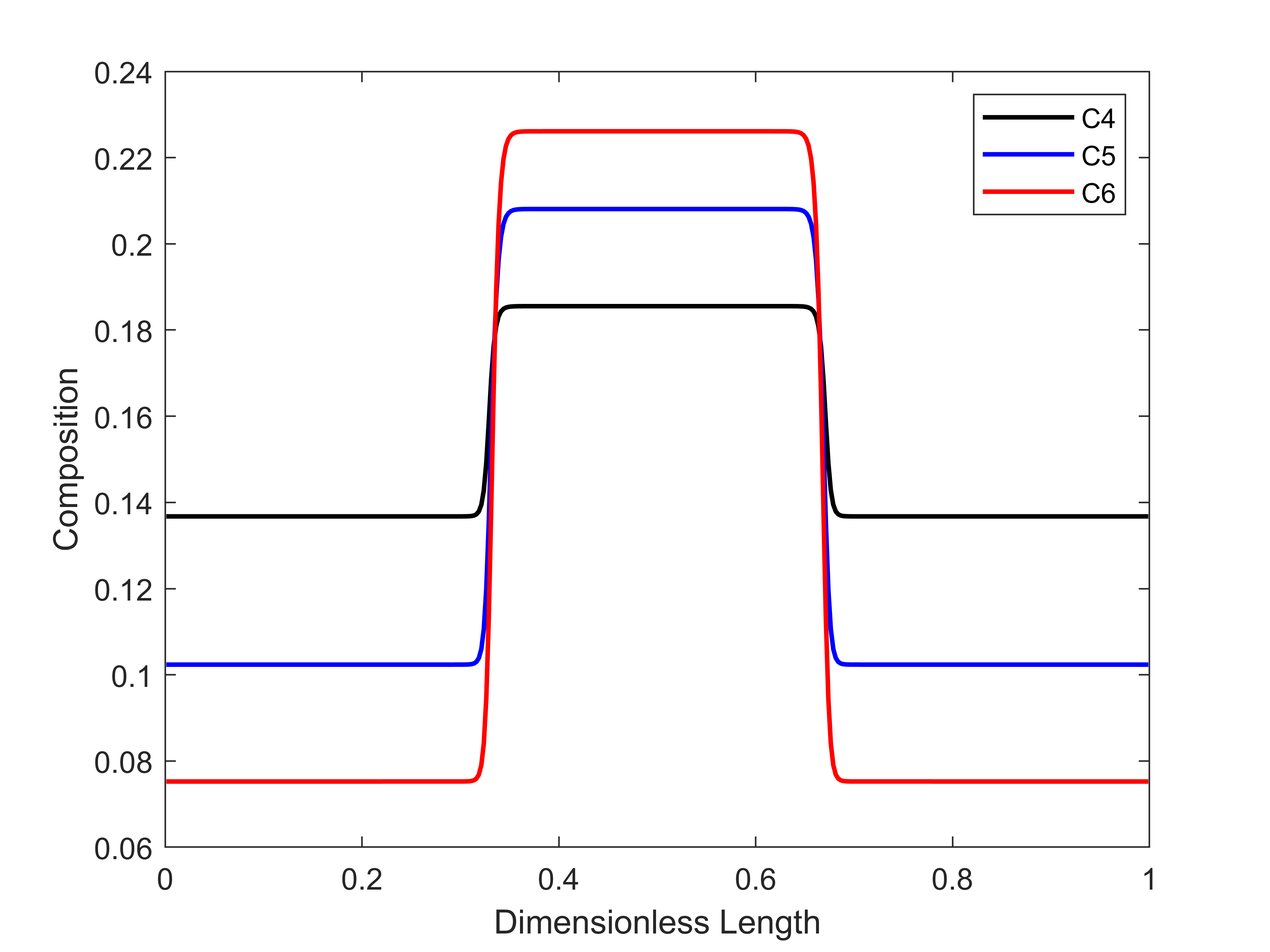}
        \end{subfigure}
        \hfill      
\caption{The equilibrium density and composition profiles obtained from LBM simulations for Case 6, described in Table \ref{tabSixComp}. (a) The density vs dimensionless length ($x/n_x$). (b) Composition of C1, C2, and C3 vs dimensionless length. (c) Composition of C4, C5, and C6 vs dimensionless length.}
\label{figSixComp}
\end{figure}

Here $S_V$ is the saturation (volume fraction) of the vapor phase. $W$ is set to be 4, and $\rho_{i, V}$, $\rho_{i, L}$, and $S_V$  are calculated by performing a flash calculation for the given mixture at the relevant $p$, $T$, and $z_i$. Each simulation is run for 1,000,000 time-steps to achieve equilibrium.
The results of the equilibrium density and composition profiles for one of the cases, Case 6, are presented in Figure \ref{figSixComp}. The equilibrium density and composition in the vapor and liquid phases for each cases is summarized in Table \ref{tab1to6Results}.

\begin{table}[H]
\caption{The density and composition in the vapor and liquid phase obtained from LBM simulations for Cases 1-6 (described in Table \ref{tabSixComp}).}
\label{tab1to6Results}
\centering
\begin{tabular}{lcccccccc}
\hline \hline
Case     & Phase    & $\rho$ ($kg/m^3$)  & $x_{C1}$  & $x_{C2}$  & $x_{C3}$   & $x_{C4}$  & $x_{C5}$  & $x_{C6}$  \\
\hline
\multirow{2}{*}{1} & Vapor  & 55.43  & 1.0000   & -        & -        & -        & -        & -        \\
                   & Liquid & 284.87 & 1.0000   & -        & -        & -        & -        & -        \\
\multirow{2}{*}{2} & Vapor  & 44.53  & 0.8538   & 0.1462   & -        & -        & -        & -        \\
                   & Liquid & 434.50 & 0.4366   & 0.5634   & -        & -        & -        & -        \\
\multirow{2}{*}{3} & Vapor  & 44.34  & 0.7527   & 0.1897   & 0.0577   & -        & -        & -        \\
                   & Liquid & 480.09 & 0.2599   & 0.3578   & 0.3822   & -        & -        & -        \\
\multirow{2}{*}{4} & Vapor  & 53.05  & 0.5639   & 0.2599   & 0.1225   & 0.0536   & -        & -        \\
                   & Liquid & 475.87 & 0.1567   & 0.2462   & 0.2877   & 0.3094   & -        & -        \\
\multirow{2}{*}{5} & Vapor  & 67.53  & 0.4050   & 0.2625   & 0.1700   & 0.1024   & 0.0601   & -        \\
                   & Liquid & 448.13 & 0.1121   & 0.1725   & 0.2123   & 0.2420   & 0.2611   & -        \\
\multirow{2}{*}{6} & Vapor  & 90.08  & 0.2831   & 0.2240   & 0.1786   & 0.1368   & 0.1024   & 0.0752   \\
                   & Liquid & 401.37 & 0.0923   & 0.1295   & 0.1584   & 0.1855   & 0.2081   & 0.2261   \\
\hline \hline
\end{tabular}
\end{table}

To ensure that the results presented in Table \ref{tab1to6Results} are consistent with thermodynamics, and do not violate the iso-fugacity criterion (or equivalently the iso-chemical potential criterion), they are compared to the theoretical predictions obtained from a flash calculation. The flash calculations are performed at the initial temperature and overall compositions for each of the cases. However, in the LBM simulation, the pressure deviates from its initial value as Equation \ref{eqInit1D} represent only an approximation to the equilibrium profile. The theoretical flash calculations were therefore performed using the updated pressure values obtained from the LBM simulations. The relative errors between the LBM simulation results and the theoretical predictions are summarized in Table \ref{tab1to6Errors}.

\begin{table}[H]
\caption{The relative error (\%) between LBM simulation results and the theoretical predictions for each of the density and composition values given in Table \ref{tab1to6Results}.}
\label{tab1to6Errors}
\centering
\begin{tabular}{lcccccccc}
\hline \hline
Case               & Phase  & $\rho$ error (\%)   &  $x_{C1}$ error (\%) & $x_{C2}$ error (\%) & $x_{C3}$ error (\%) &  $x_{C4}$ error (\%) &  $x_{C5}$ error (\%) &  $x_{C6}$ error (\%) \\
\hline
\multirow{2}{*}{1} & Vapor  & 0.000002 & 0.000000 & -        & -        & -        & -        & -        \\
                   & Liquid & 0.000001 & 0.000000 & -        & -        & -        & -        & -        \\
\multirow{2}{*}{2} & Vapor  & 0.000021 & 0.000015 & 0.000089 & -        & -        & -        & -        \\
                   & Liquid & 0.000035 & 0.000065 & 0.000050 & -        & -        & -        & -        \\
\multirow{2}{*}{3} & Vapor  & 0.022743 & 0.033667 & 0.173693 & 0.133215 & -        & -        & -        \\
                   & Liquid & 0.020379 & 0.013962 & 0.170605 & 0.150722 & -        & -        & -        \\
\multirow{2}{*}{4} & Vapor  & 0.069883 & 0.130091 & 0.263293 & 0.122784 & 0.194704 & -        & -        \\
                   & Liquid & 0.036333 & 0.086755 & 0.264003 & 0.093838 & 0.254691 & -        & -        \\
\multirow{2}{*}{5} & Vapor  & 0.093856 & 0.210405 & 0.189726 & 0.231818 & 0.062673 & 0.179850 & -        \\
                   & Liquid & 0.043724 & 0.142220 & 0.208580 & 0.215747 & 0.012199 & 0.264909 & -        \\
\multirow{2}{*}{6} & Vapor  & 0.094781 & 0.232504 & 0.074052 & 0.186282 & 0.159989 & 0.040750 & 0.137930 \\
                   & Liquid & 0.046245 & 0.148373 & 0.113146 & 0.192296 & 0.132857 & 0.018878 & 0.231750 \\
\hline \hline
\end{tabular}
\end{table}

\begin{figure}[H]
    \centering
    \includegraphics[scale=0.65]{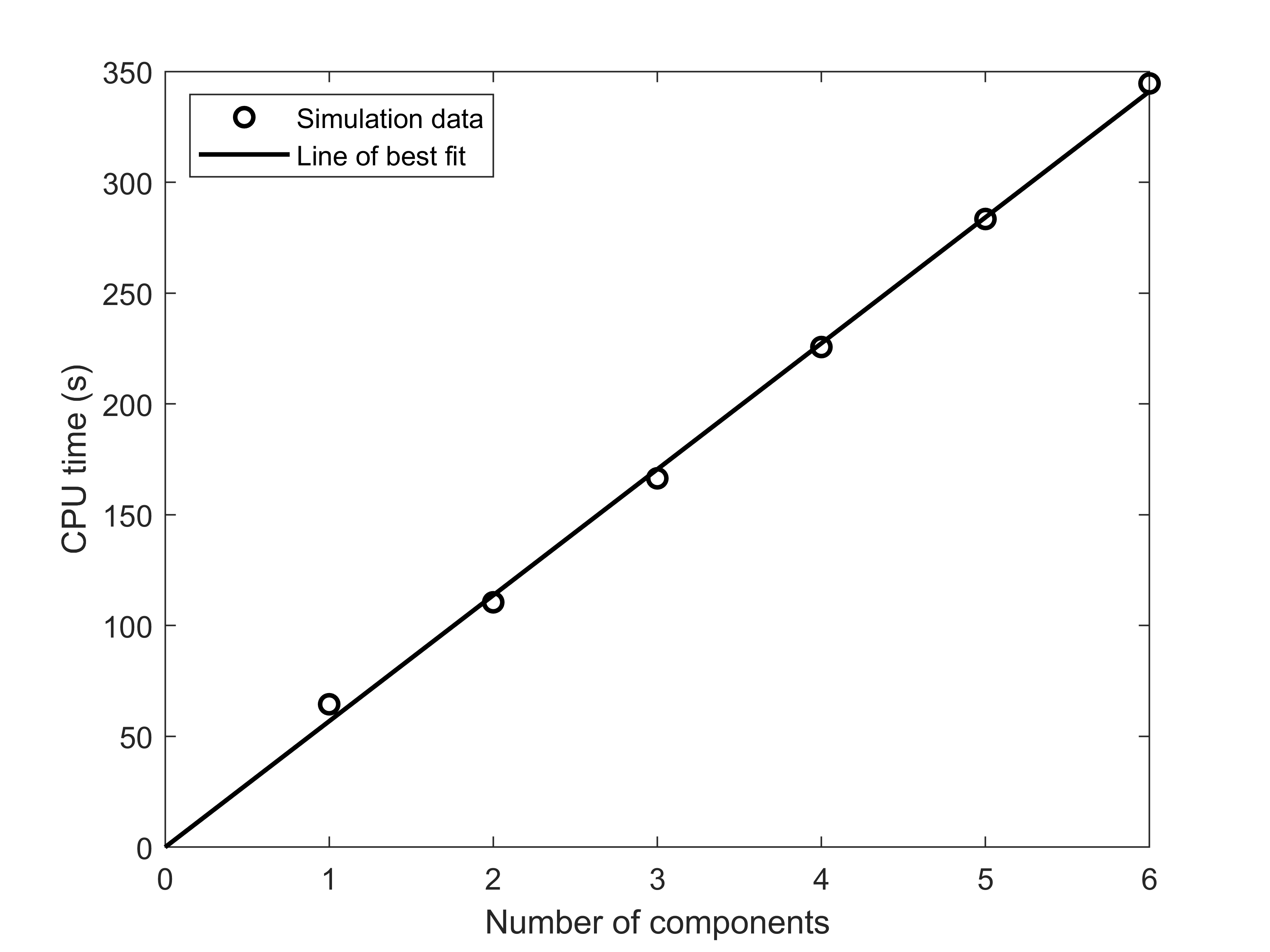}
    \caption{CPU time vs number of components. The dots represent the data from simulations and the solid line is the line of best fit.}
    \label{figCPUTime}
\end{figure}

Table \ref{tab1to6Errors} shows the maximum error to be 0.264909\% which indicates excellent agreement with thermodynamic predictions. It should be noted that
in these (small) maximum errors are typically associated with the compositions of the chemical components present in the smallest relative quantities for which computational round-off errors are prone to be more significant. This study offers the opportunity to quantitatively test how the computational time in a partially-miscible LBM simulation scales with the number of components used. We plotted the CPU time for each simulation case against the number of components used, and the results are presented in Figure \ref{figCPUTime}.

Figure \ref{figCPUTime} demonstrates that the computational time required for an LBM simulation scales linearly with the number of components present in the system. This is a notable difference from conventional thermodynamic flash calculations, where computational times increase exponentially with the number of components in the system \cite{Li2006}. 

\subsection{Ternary vapor-liquid equilibrium}
\label{secTernaryVLE}

Section \ref{secMulticomp} presented multicomponent simulations at a single condition for each mixture. However, in this section, we present a more detailed investigation of ternary systems at a broad range of conditions, and provide a comprehensive analysis of their phase behavior. To the best of our knowledge, such a thorough analysis of ternary phase behavior is currently absent in the literature. 
We perform a series of simulations of VLE with a flat interface to obtain the compositions of the vapor and liquid phases at different pressure, temperature, and overall composition conditions. The results are used to generate different characteristic ternary diagrams and their accuracy is tested by comparing them with the results predicted by a flash calculation.

\begin{figure}[H]
    \centering
        \begin{subfigure}{0.49\textwidth}
            \centering
            \caption{}
            \includegraphics[width=\textwidth]{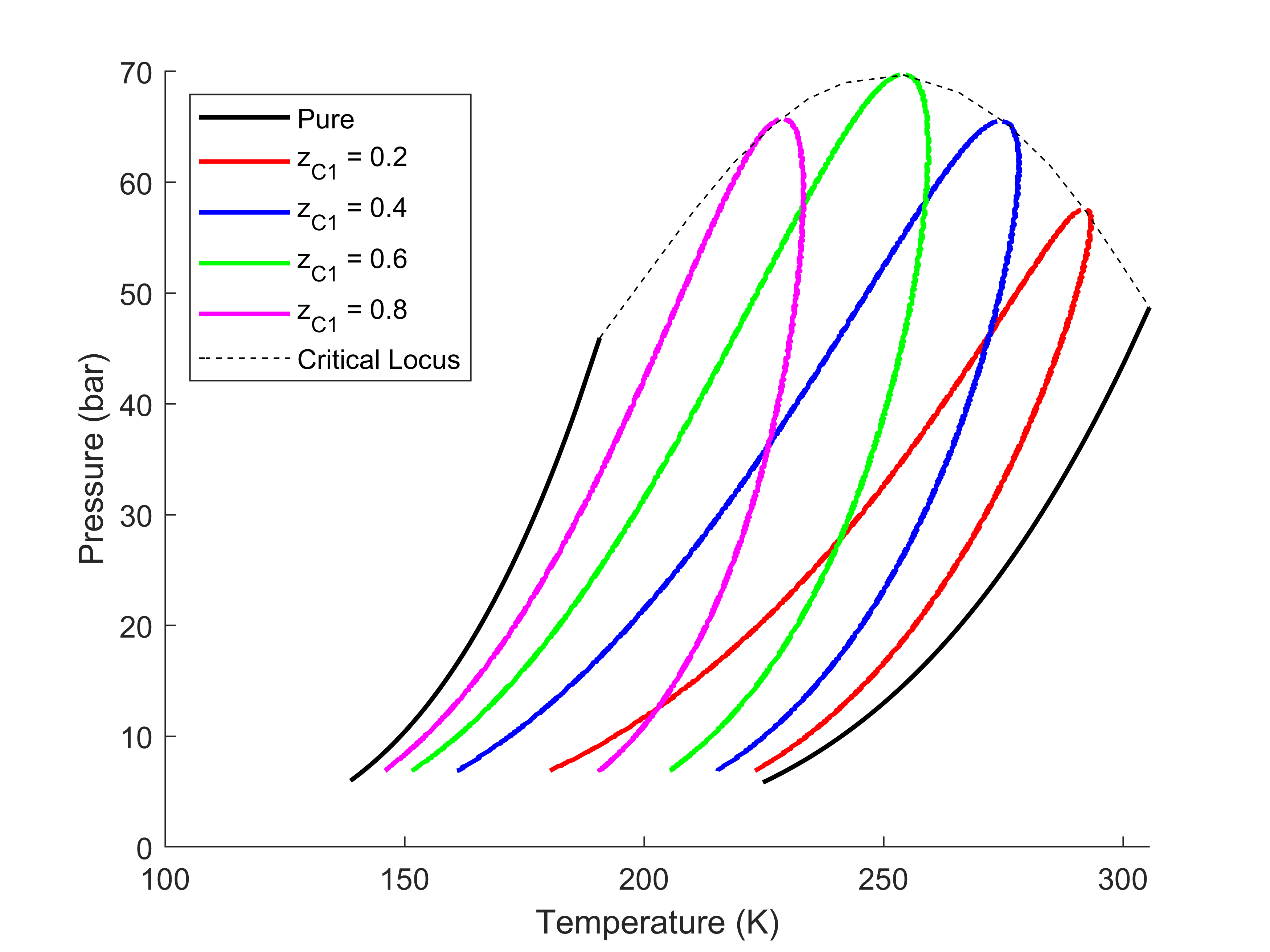}
        \end{subfigure}%
        \hfill
        \begin{subfigure}{0.49\textwidth}
            \centering
            \caption{}
            \includegraphics[width=\textwidth]{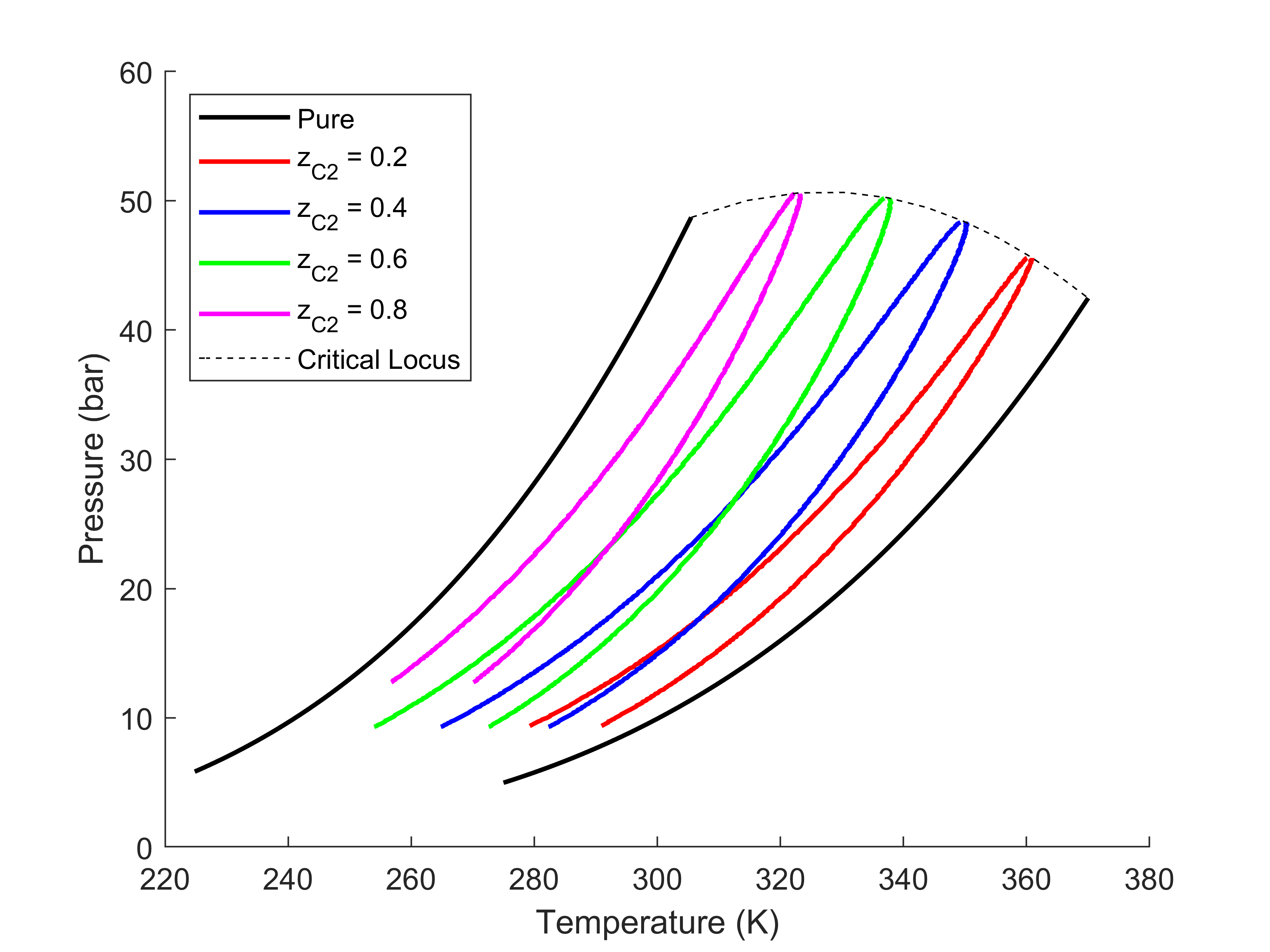}
        \end{subfigure}
        \hfill        
        \begin{subfigure}{0.49\textwidth}
            \centering
            \caption{}
            \includegraphics[width=\textwidth]{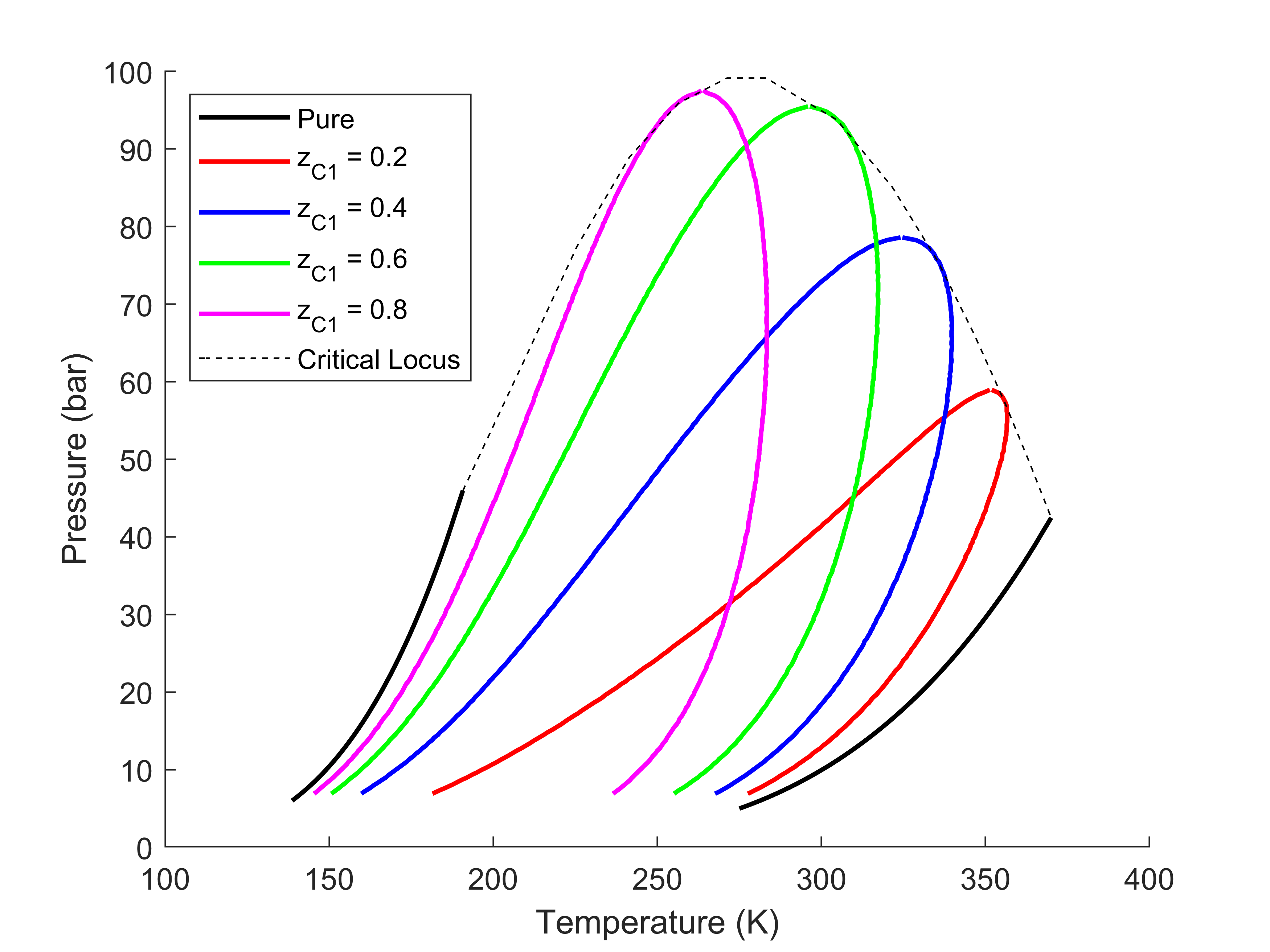}
        \end{subfigure}
        \hfill 

\caption{The theoretical pressure-temperature envelopes for a system of (a) C1-C2, (b) C2-C3, and (c) C1-C3 generated using the PR EOS at different overall compositions. The solid black lines represent the vapor-liquid equilibrium curves of pure species with the curve of the more volatile specie on the left and the less volatile specie on the right of each plot.}
\label{figPTEnvelopes}
\end{figure}

We analyze a system of C1, C2, and C3. The binary interaction parameter between each component pair is 0. 
The theoretical binary phase behavior of the C1-C2 pair, C2-C3 pair, and C1-C3 pair on the pressure-temperature plane generated using the PR EOS is shown in Figure \ref{figPTEnvelopes}.

\begin{figure}[H]
    \centering
    \includegraphics[scale=0.7]{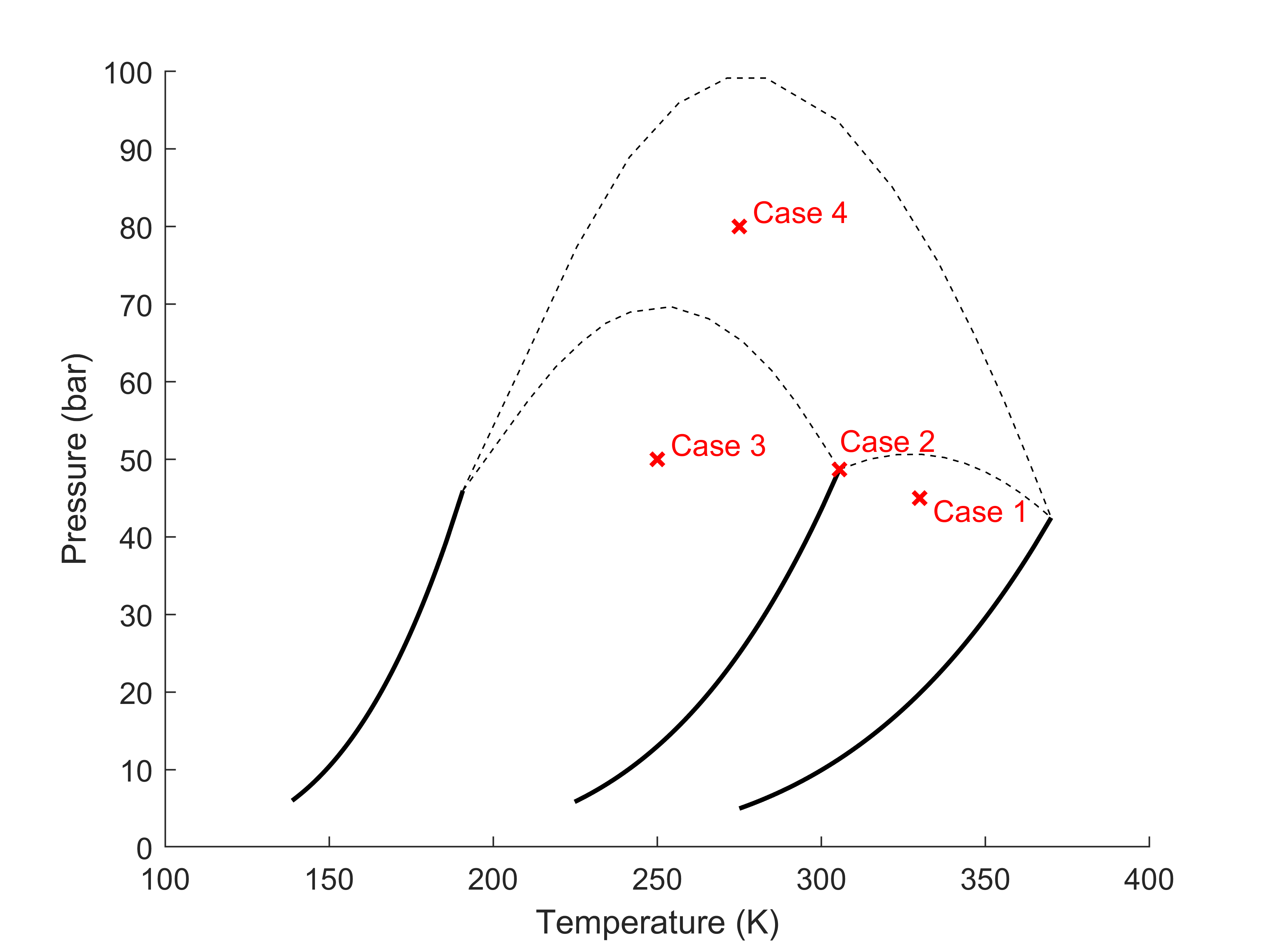}
    \caption{The pure component vapor liquid equilibrium curves for C1, C2, and C3 along with the critical loci for the C1-C2, C2-C3, and C1-C3 component pairs from Figure \ref{figPTEnvelopes}. Four cases of interest to be tested using LBM are identified in this diagram.}
    \label{figPTCases}
\end{figure}

In Figure \ref{figPTEnvelopes}, for each of the pressure-temperature graphs, the region bounded by the pure component vapor-liquid equilibrium curves and the critical locus for that component pair represents all the possible pressure-temperature conditions where two phases can coexist for that component pair. Superimposing the possible two-phase regions for all three component pairs, we get Figure \ref{figPTCases}.

\begin{figure}[H]
    \centering
        \begin{subfigure}{0.48\textwidth}
            \centering
            \caption{}
            \includegraphics[width=\textwidth]{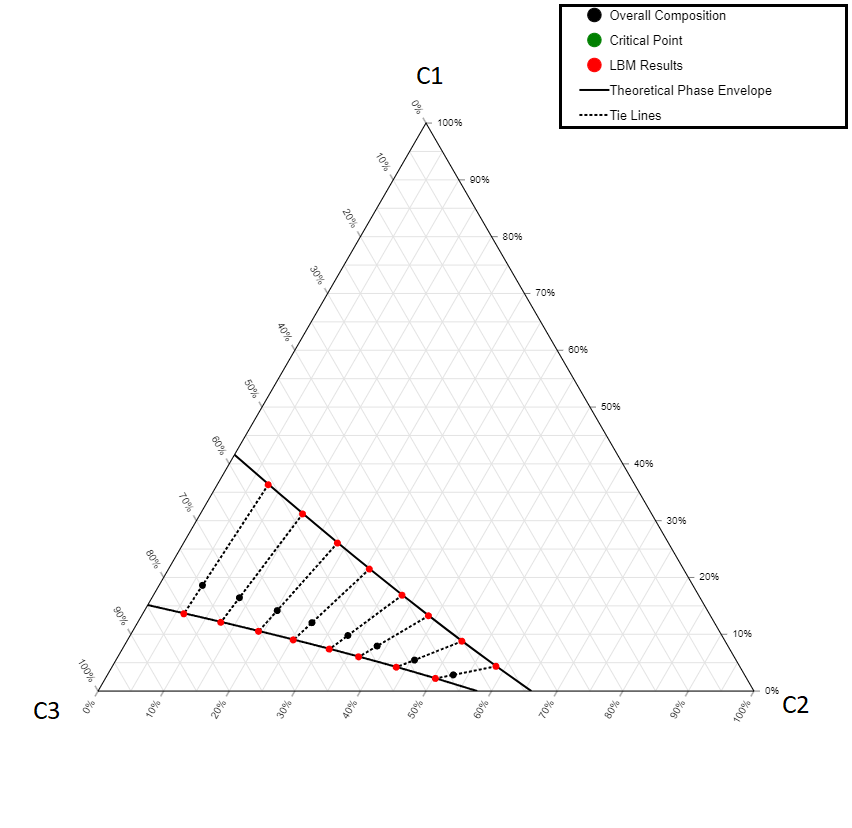}
        \end{subfigure}%
        \hfill
        \begin{subfigure}{0.48\textwidth}
            \centering
            \caption{}
            \includegraphics[width=\textwidth]{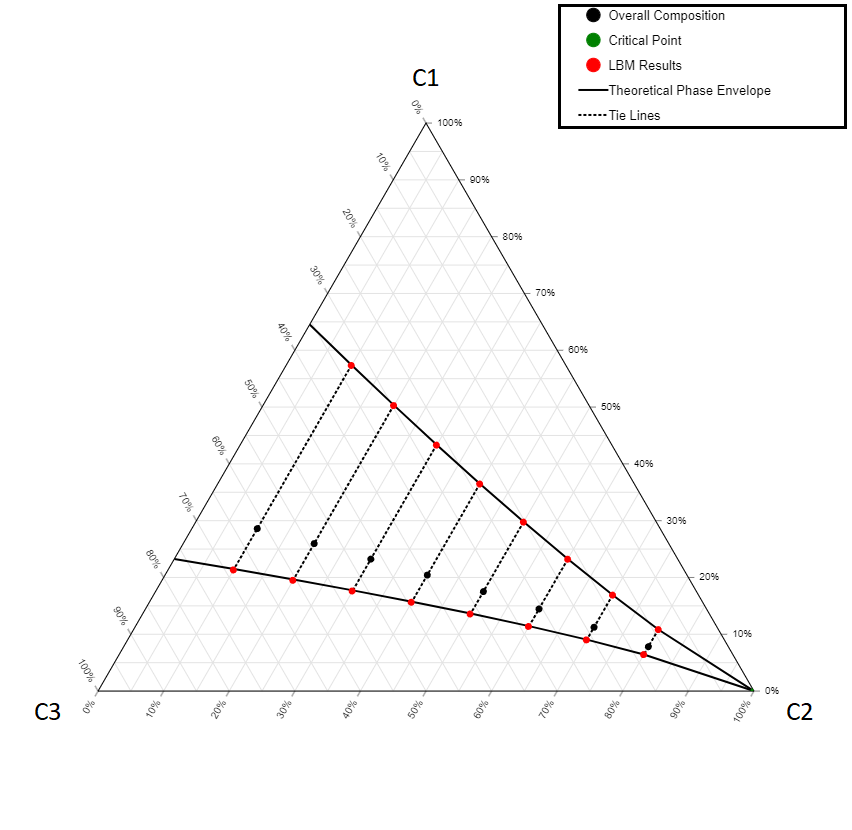}
        \end{subfigure}
        \hfill        
        \begin{subfigure}{0.48\textwidth}
            \centering
            \caption{}
            \includegraphics[width=\textwidth]{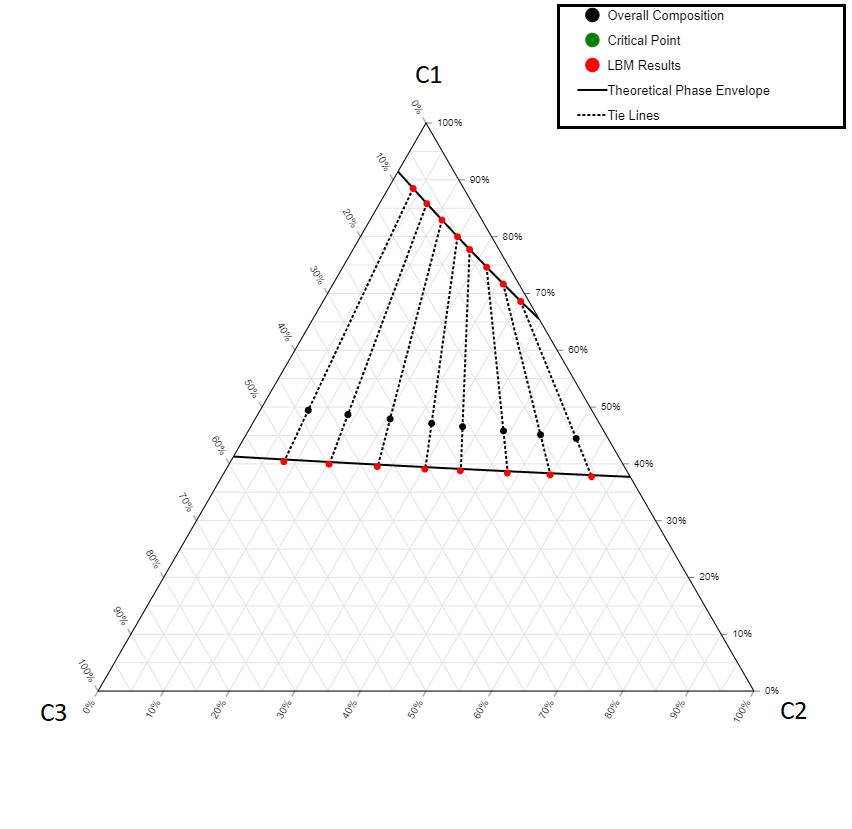}
        \end{subfigure}
        \hfill 
        \begin{subfigure}{0.48\textwidth}
            \centering
            \caption{}
            \includegraphics[width=\textwidth]{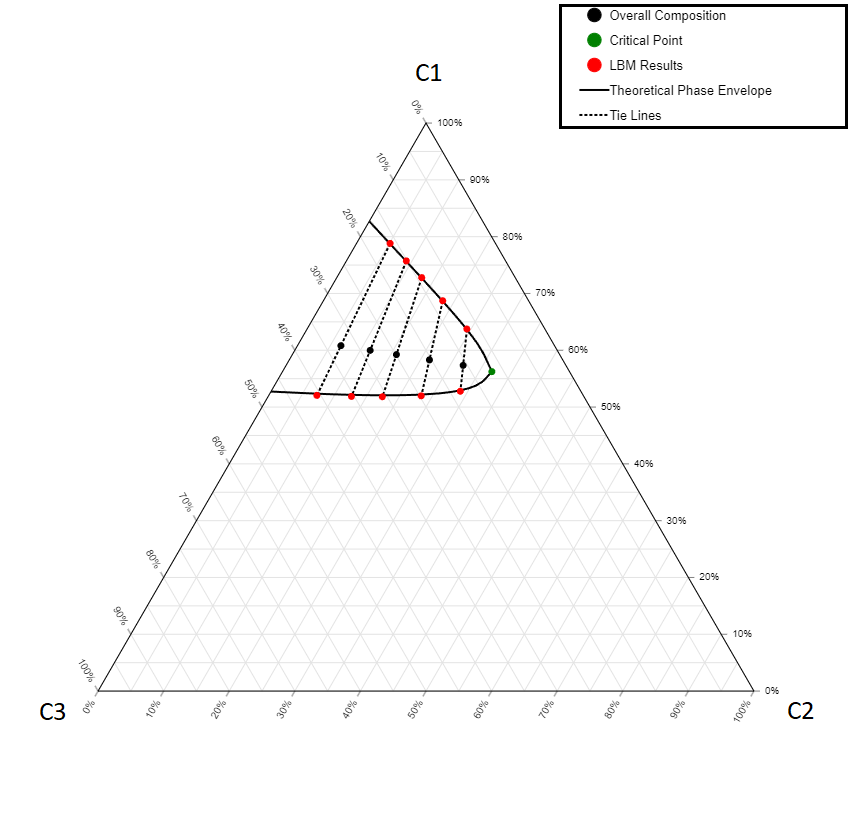}
        \end{subfigure}
        \hfill 

\caption{Ternary diagram for (a) Case 1, (b) Case 2, (c) Case 3, and (d) Case 4. The solid black lines connect the theoretical compositions from the vapor phase and liquid phase obtained through flash calculations using the PR EOS and they form the theoretical phase envelope. The solid black dots represent the overall composition at which each LBM simulation is carried out, with the dashed black lines representing the theoretical tie lines at those overall compositions. The red dots represent the compositions of the vapor and liquid phase obtained from the LBM simulations at equilibrium. The green dot represents the critical point (where applicable).}
\label{figTernaryPlots}
\end{figure}

Four pressure-temperature cases of interest are identified in Figure \ref{figPTCases}, which will be tested using the LBM, and the resulting phase compositions will be plotted on a ternary diagram. In case 1 ($p=45\ bar$ and $T=330\ K$), we should see two phases for the C2-C3 side of the triangle as well as the C1-C3 side. In case 2 ($p=48.71\ bar$ and $T=305.51\ K$), two phases exist for the pure C2 corner of the triangle and C1-C3 side. In case 3 ($p=50\ bar$ and $T=250\ K$), two phases exist for the C1-C2 side and C1-C3 side. In case 4 ($p=80\ bar$ and $T=275\ K$), two phases exist for only the C1-C3 side. These four cases are tested in LBM at different overall compositions to generate the ternary diagrams of interest. For all simulations, the relevant conversions between lattice units and physical units are established by fixing the universal gas constant and the attraction parameter, co-volume, and molar mass for C1 to the following values in lattice units: $R=1$, $a_{C1}=2/49$, $b_{C1}=2/21$, and $M_{C1}=1$. The relaxation time $\tau=1.0$, and the interfacial tension parameter for the reference component, chosen to be C1, is $\kappa_i^{ref}=0.02$. The size of the computational domain is $400\times 2$ ($n_x\times n_y$), and the density of each component along the $x$ direction is initialized using Equation \ref{eqInit1D}, with $W$ set to be 4, and $\rho_{i, V}$, $\rho_{i, L}$, and $S_V$ obtained by performing a flash calculation for the given mixture at the desired pressure, temperature and overall composition. To achieve equilibrium, the simulations are run for 1,000,000 time steps. The vapor and liquid phase compositions obtained at different overall compositions, for each of the four pressure-temperature cases are plotted on the ternary diagrams shown in Figure \ref{figTernaryPlots}. The theoretical phase envelopes and tie lines for the respective overall composition of each of the simulations are also shown.

It can be seen that the LBM correctly predicts the phase behavior of ternary systems at a multitude of different pressure, temperature and composition conditions. This clearly showcases the capability of the fugacity-based LBM in capturing partially-miscible phases. The proposed LBM model accurately captures the full thermodynamic behavior of ternary systems, which in the past had been mostly constrained to immiscible applications.

\subsection{Three-component three-phase case}
\label{secThreePhase}

Until now, our results have focused on two-phase systems. In this section, we extend our simulations to the case of three-phase equilibrium, with flat interfaces between phases. To achieve this, we employ a mixture of C1, C4, and CO2, with the properties of these components given in Table \ref{tabCompPropThreePhase}. The properties listed in Table \ref{tabCompPropThreePhase} differ slightly from those in Table \ref{tabCompProp}, as we utilized an open-source code for three-phase flash calculations in this section (the code can be found as a supplementary material to Ref. \cite{Michelsen2007}). The three-phase flash code employs fixed component properties that differ from those used in earlier (and following) sections of this study.

\begin{table}[h]
\caption{The properties of relevant components used in the LBM simulations in Section \ref{secThreePhase}.}
\label{tabCompPropThreePhase}
\centering
\begin{tabular}{lcccc}
\hline \hline
Component & Critical Pressure (bar)\ \  & Critical Temperature (K)\ \  & Acentric factor\ \  & Molar Mass (g/mol)\ \  \\
\hline
Carbon   Dioxide (CO2) & 73.75 & 304.20 & 0.2250 & 44.010  \\
Methane (C1)           & 45.99 & 190.60 & 0.0080 & 16.043  \\
n-Butane (C4)          & 37.99 & 425.20 & 0.1930 & 58.123  \\

\hline \hline
\end{tabular}
\end{table}

\begin{figure}[H]
    \centering
        \begin{subfigure}{0.49\textwidth}
            \centering
            \caption{}
            \includegraphics[width=\textwidth]{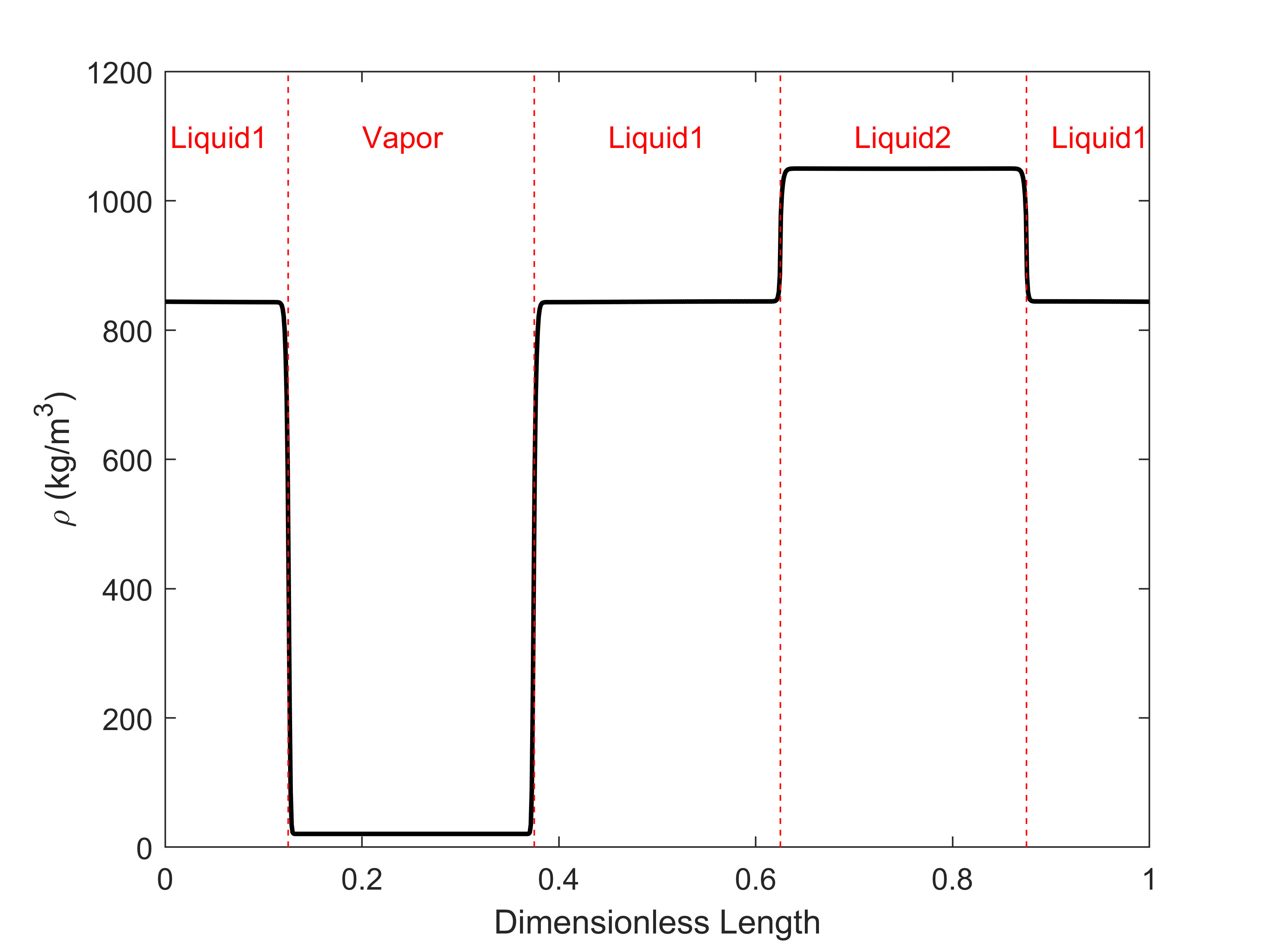}
        \end{subfigure}%
        \hfill
        \begin{subfigure}{0.49\textwidth}
            \centering
            \caption{}
            \includegraphics[width=\textwidth]{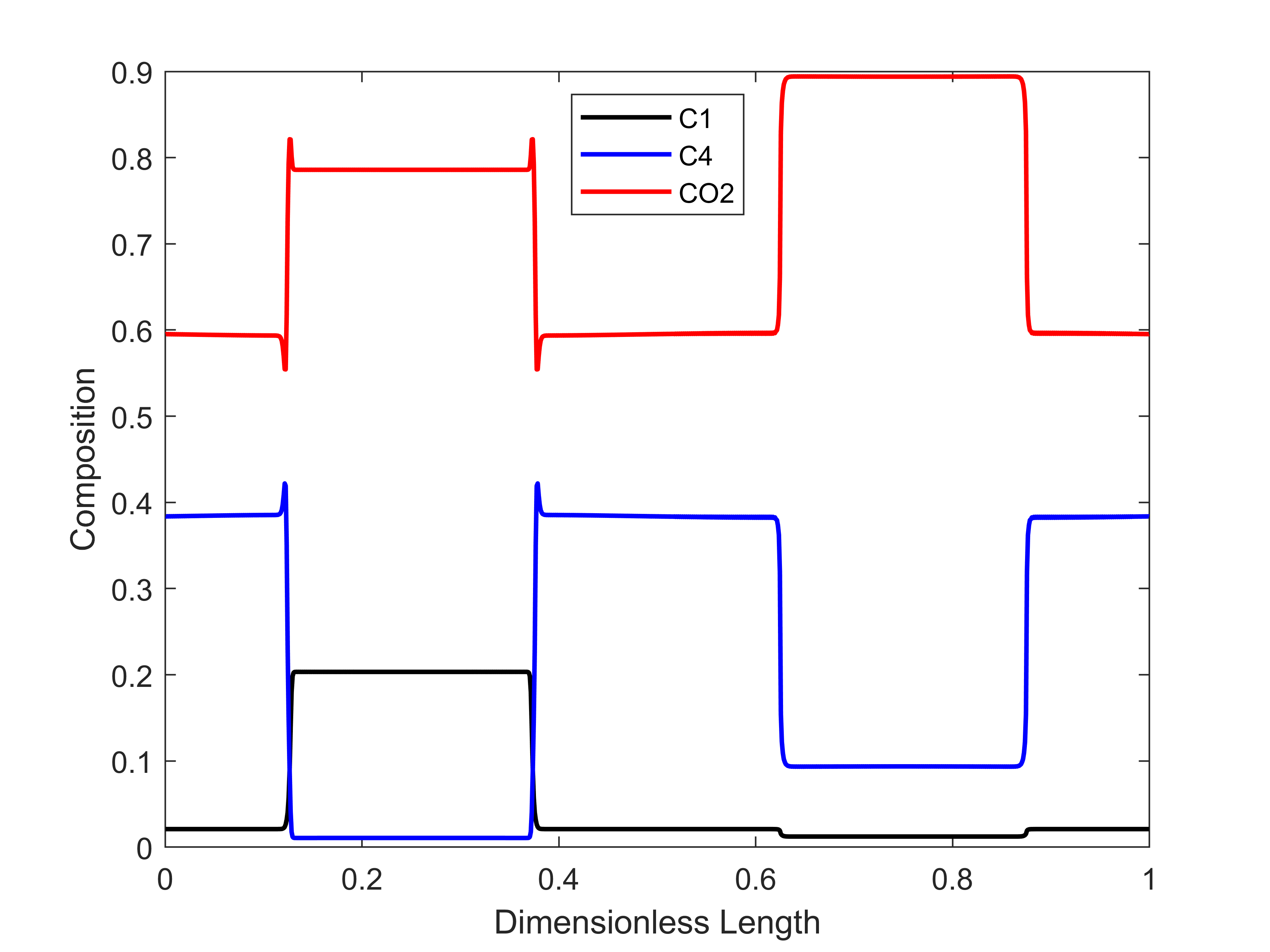}
        \end{subfigure}
        \hfill        
  
\caption{The equilibrium density and composition profile obtained from the LBM simulation. (a) Density vs dimensionless length ($x/n_x$). (b) Composition of C1, C4, and CO2 vs dimensionless length. The three distinct phases (vapor, liquid1, and liquid2) are marked on (a).}
\label{figTheePhasePlots}
\end{figure}

The binary interaction parameters for the CO2-C1 and CO2-C4 pairs are 0.12 and 0.15, respectively, while the remaining binary interaction parameters are all zero. The simulation is conducted at a temperature of 225 K, an initial pressure of 9 bar, and an overall composition of $z_{C1}=0.0194$, $z_{C4}=0.2643$, and $z_{CO2}=0.7163$. Under these conditions, this mixture will form three phases: vapor, liquid1, and liquid2 (in order of increasing density).
The relevant conversions between lattice units and physical units are established by fixing the universal gas constant and the attraction parameter, co-volume, and molar mass for C1 to the following values in lattice units: $R=1$, $a_{C1}=0.011550$, $b_{C1}=0.107781$, and $M_{C1}=1$. The relaxation time $\tau=1.0$, and the interfacial tension parameter for the reference component, chosen to be C1, is $\kappa_i^{ref}=0.02$. The size of the computational domain is $1000\times 2$ ($n_x\times n_y$), and the density of each component along the $x$ direction is initialized as shown in Equation \ref{eqInit3Phase} (the domain will be symmetric in the $y$ direction),

\begin{equation}
    \label{eqInit3Phase}
    \rho_i(x,y,t=0)=
    \begin{cases}
    \dfrac{\rho_{i,L1}+\rho_{i,V}}{2}-\dfrac{\rho_{i,L1}-\rho_{i,V}}{2}\tanh{\left[\dfrac{2\left(x-S1\right)}{W}\right]} & \text{if $0 < x \leq \dfrac{n_x}{4}$} \\ 
    \dfrac{\rho_{i,V}+\rho_{i,L1}}{2}-\dfrac{\rho_{i,V}-\rho_{i,L1}}{2}\tanh{\left[\dfrac{2\left(x-S2\right)}{W}\right]} & \text{if $\dfrac{n_x}{4} < x \leq \dfrac{n_x}{2}$} \\
    \dfrac{\rho_{i,L1}+\rho_{i,L2}}{2}-\dfrac{\rho_{i,L1}-\rho_{i,L2}}{2}\tanh{\left[\dfrac{2\left(x-S3\right)}{W}\right]} & \text{if $\dfrac{n_x}{2} < x \leq \dfrac{3}{4}n_x$} \\
    \dfrac{\rho_{i,L2}+\rho_{i,L1}}{2}-\dfrac{\rho_{i,L2}-\rho_{i,L1}}{2}\tanh{\left[\dfrac{2\left(x-S1\right)}{W}\right]} & \text{if $\dfrac{3}{4}n_x < x \leq n_x$} \\
    \end{cases},
\end{equation}

where $S1$, $S2$, $S3$, and $S4$ are $n_x/8$, $3n_x/8$, $5n_x/8$, and $7n_x/8$, respectively. This is because, at the given initial temperature, pressure, and overall composition, the saturation (volume fraction) of the vapor phase, liquid1 phase, and liquid2 phase is 0.25, 0.5, and 0.25, respectively. The interface width is $W=8$. $\rho_{i, V}$, $\rho_{i, L1}$, and $\rho_{i, L2}$ are the densities of component `$i$' in the vapor phase, liquid1 phase, and liquid2 phase, respectively, obtained through a flash calculation at the initial temperature, pressure, and overall composition. 
The simulation is run for 10,000,000 time-steps to achieve equilibrium. The results of the equilibrium density and composition profiles are presented in Figure \ref{figTheePhasePlots}.

To test whether the results presented in Figure \ref{figTheePhasePlots} are consistent with thermodynamic predictions, we compare them with the results of a three-phase flash calculation performed using the temperature, pressure, and overall composition from the LBM equilibrium profiles. It should be noted that the LBM temperature and overall composition will remain constant at their initial values, whereas the pressure will slightly change, as discussed in Section \ref{secMulticomp}. The resulting relative errors between the values obtained from the LBM simulation and theoretical values from the flash calculation are summarized in Table \ref{tabThreephaseError}.

\begin{table}[H]
\caption{The relative error (\%) between LBM simulation results and the theoretical predictions. These errors are reported for the density and composition for each phase.}
\label{tabThreephaseError}
\centering
\begin{tabular}{lcccc}
\hline \hline
Phase   & $\rho$ error (\%) & $x_{C1}$ error (\%) & $x_{C4}$ error (\%) & $x_{CO2}$ error (\%) \\
\hline
Vapor   & 0.002536          & 0.247283            & 0.056008            & 0.064564             \\
Liquid1 & 0.015068          & 0.273085            & 0.063098            & 0.050260             \\
Liquid2 & 0.000392          & 0.235150            & 0.008564            & 0.002343             \\
\hline \hline
\end{tabular}
\end{table}

The low values of relative errors reported in Table \ref{tabThreephaseError} indicate excellent agreement between the results obtained from the LBM simulation and theoretical predictions, demonstrating that our methodology is not limited to two-phase systems but is generalizable to any number of phases.

\subsection{Ten-component hydrocarbon mixture}
\label{sec10Comp}

In this section, we demonstrate the ability of our model to handle a large number of components by simulating a 10-component hydrocarbon mixture taken from Ref. \cite{Kenyon1987}. The mixture consists of all the components listed in Table \ref{tabCompProp}, and its overall composition is presented in Table \ref{tabCompzi}. The binary interaction parameters ($\delta_{ij}$) between all component pairs are zero, except for the pairs that involve CO2. The values of $\delta_{ij}$ for each component with CO2 are also provided in Table \ref{tabCompzi}.

\begin{table}[H]
\caption{The overall composition of the ten-component mixture and the binary interaction parameter of each of the components with CO2.}
\label{tabCompzi}
\centering
\begin{tabular}{lcc}
\hline \hline
Component & Overall Composition\ \ & $\delta_{ij}$ with CO2  \\
\hline
Carbon   Dioxide (CO2) & 0.0031  & 0.000 \\
Methane (C1)           & 0.6192  & 0.105 \\
Ethane (C2)            & 0.1408  & 0.130 \\
Propane (C3)           & 0.0835  & 0.125 \\
iso-Butane (iC4)       & 0.0097  & 0.120 \\
n-Butane (C4)          & 0.0341  & 0.115 \\
iso-Pentane (iC5)      & 0.0084  & 0.115 \\
n-Pentane (C5)         & 0.0148  & 0.115 \\
n-Hexane (C6)          & 0.0179  & 0.115 \\
C7+                    & 0.0685  & 0.115 \\
\hline \hline
\end{tabular}
\end{table}

\begin{figure}[H]
    \centering
        \begin{subfigure}{0.49\textwidth}
            \centering
            \caption{}
            \includegraphics[width=\textwidth]{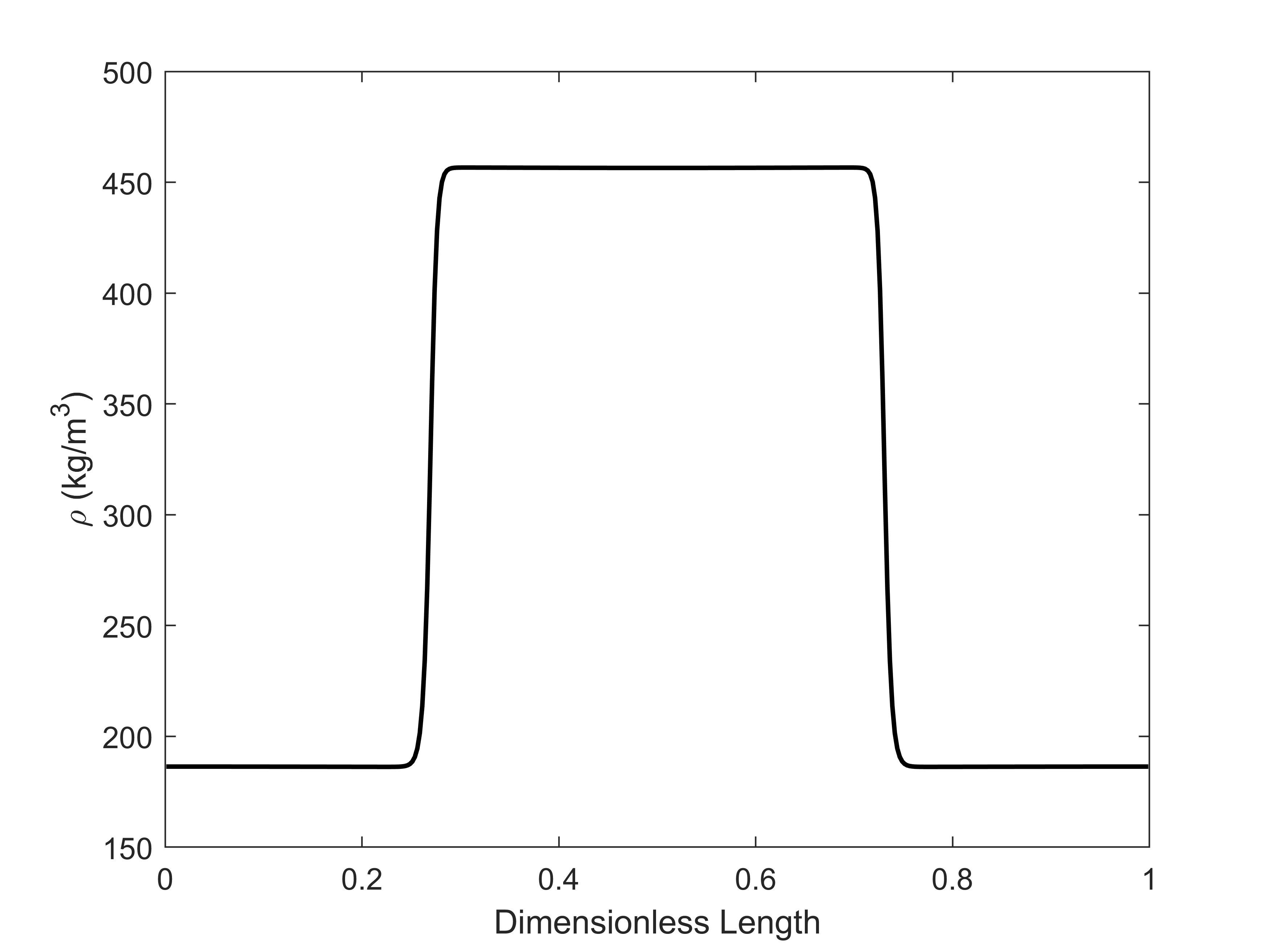}
        \end{subfigure}%
        \hfill
        \begin{subfigure}{0.49\textwidth}
            \centering
            \caption{}
            \includegraphics[width=\textwidth]{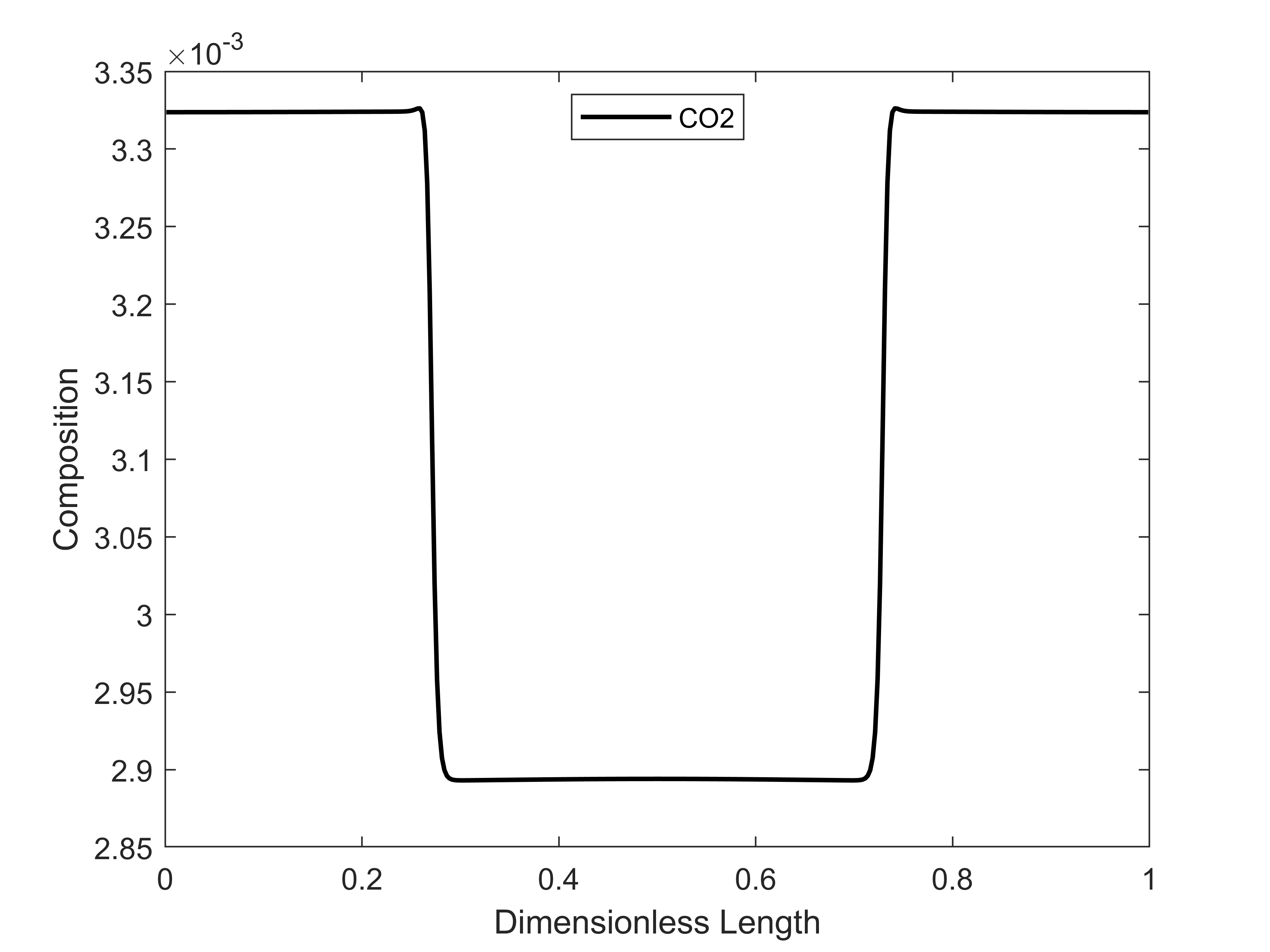}
        \end{subfigure}
        \hfill        
        \begin{subfigure}{0.49\textwidth}
            \centering
            \caption{}
            \includegraphics[width=\textwidth]{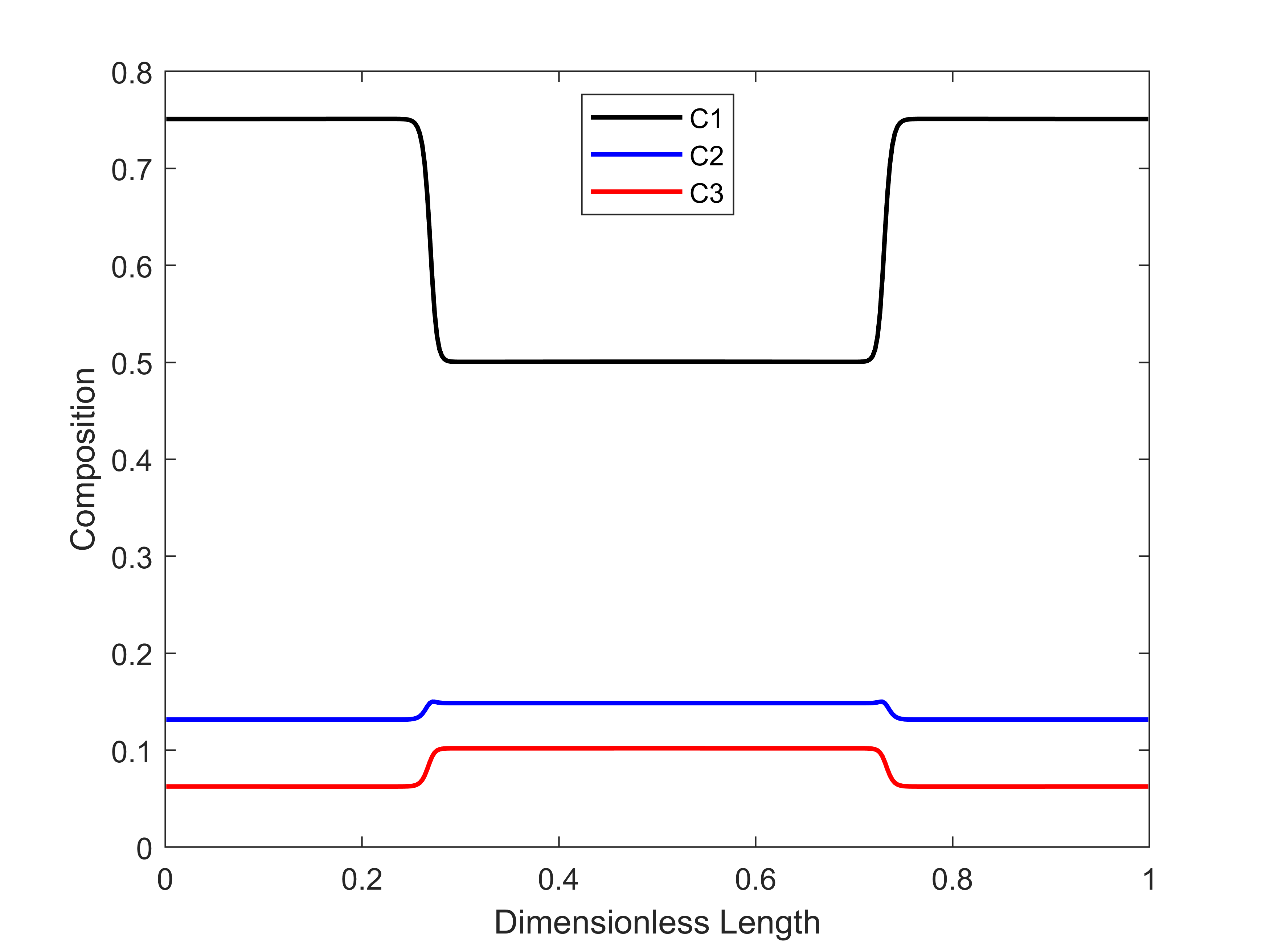}
        \end{subfigure}
        \hfill 
                \begin{subfigure}{0.49\textwidth}
            \centering
            \caption{}
            \includegraphics[width=\textwidth]{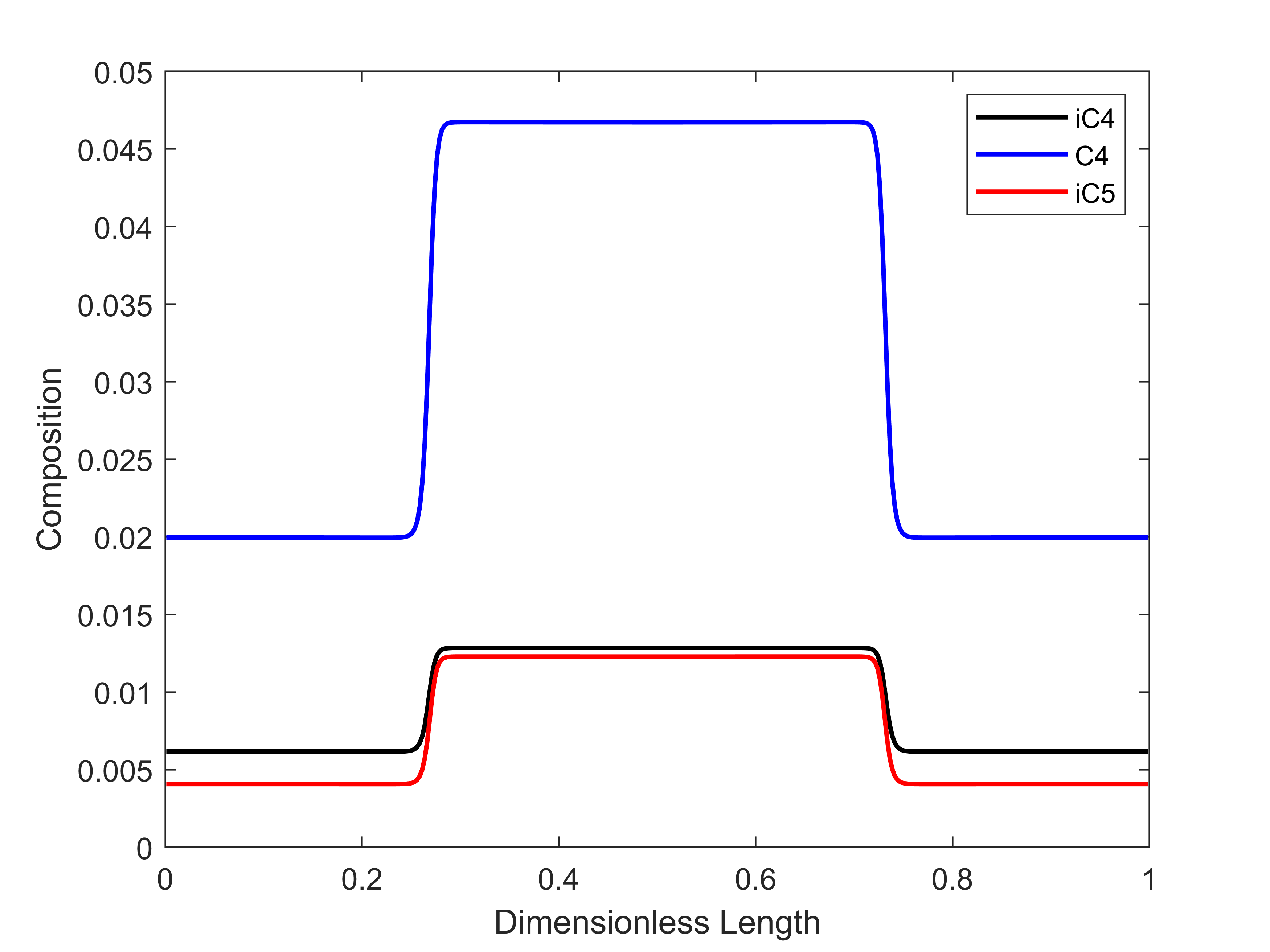}
        \end{subfigure}
        \hfill 
                \begin{subfigure}{0.49\textwidth}
            \centering
            \caption{}
            \includegraphics[width=\textwidth]{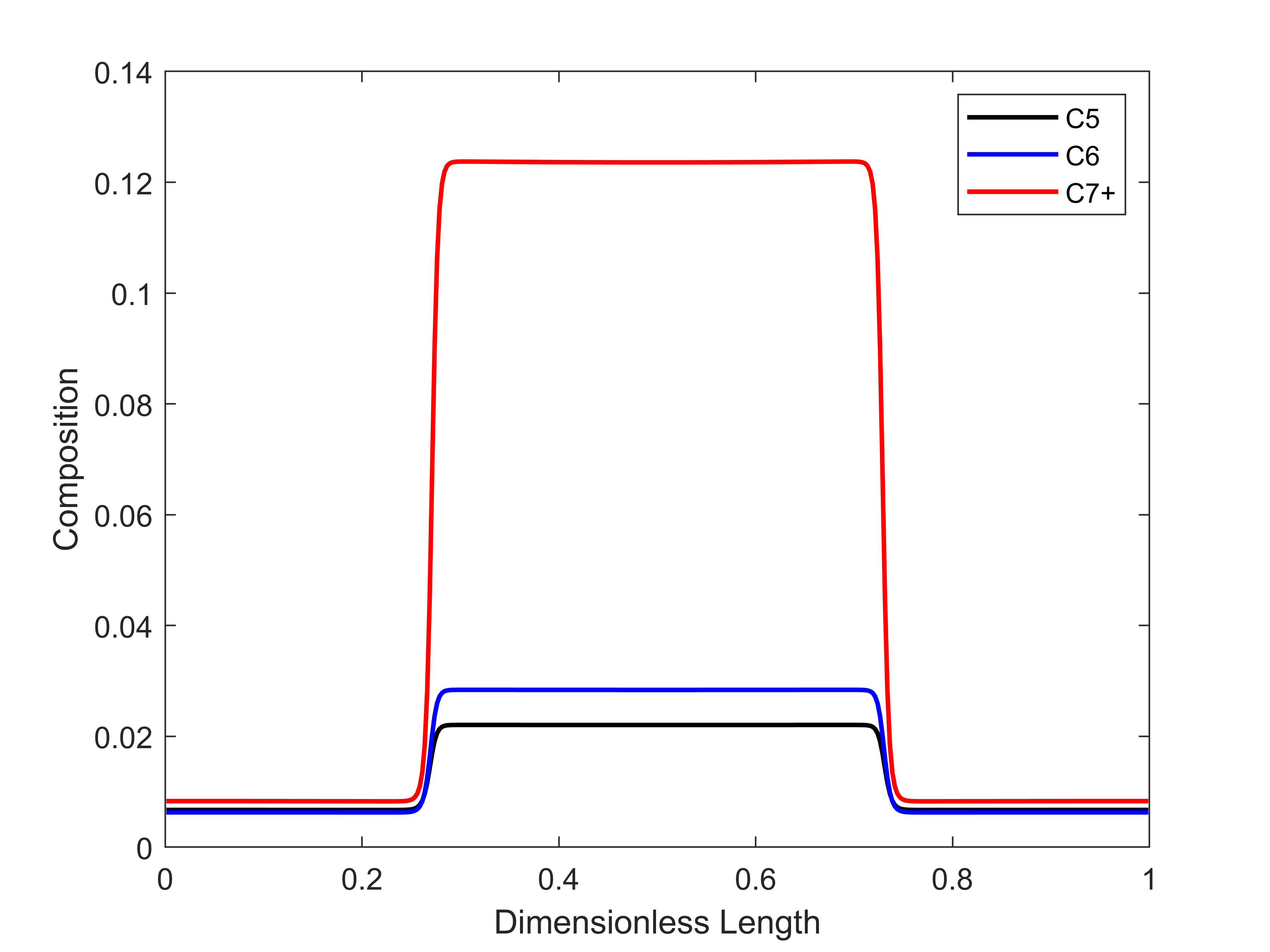}
        \end{subfigure}
        \hfill 

\caption{The equilibrium density and composition profiles obtained from LBM simulations for the ten-component mixture with a flat interface. (a) Density vs dimensionless length ($x/n_x$). (b) Composition of CO2 vs dimensionless length. (c) Composition of C1, C2, and C3 vs dimensionless length. (d) Composition of iC4, C4, and iC5 vs dimensionless length. (e) Composition of C5, C6, and C7+ vs dimensionless length.}
\label{fig10Cmpnt1D}
\end{figure}

We start off by simulating a flat interface VLE case. The mixture is initialized at a temperature of 325 K and pressure of 150 bar. The relevant conversions between lattice units and physical units are established by fixing the universal gas constant and the attraction parameter, co-volume, and molar mass for C1 to the following values in lattice units: $R=1$, $a_{C1}=2/49$, $b_{C1}=2/21$, and $M_{C1}=1$. The relaxation time $\tau=1.0$, and the interfacial tension parameter for the reference component, chosen to be C1, is $\kappa_i^{ref}=0.02$. The size of the computational domain is $400\times 2$ ($n_x\times n_y$), and the density of each component is initialized as shown in Equation \ref{eqInit1D}. $\rho_{i, V}$, $\rho_{i, L}$, and $S_V$ in Equation \ref{eqInit1D} are calculated by performing a flash calculation for the mixture at the initial $p$, $T$, and $z_i$. $W$ is set to be 4. The simulation is run for 1,000,000 time-steps and the results of the equilibrium density and composition profiles are presented in Figure \ref{fig10Cmpnt1D}.

Next, we compare the LBM results with the results of a flash calculation performed at the same conditions. The resulting relative errors between the values obtained from the LBM simulation and theoretical values from the flash calculation are summarized in Table \ref{tab10Cmpnt1DError}.

\begin{table}[H]
\caption{The relative error (\%) in the density and composition values obtained from the LBM simulation when compared with the theoretical predictions.}
\label{tab10Cmpnt1DError}
\centering
\begin{tabular}{lcc}
\hline \hline
Property  \ \    & Vapor Phase Error (\%) \ \ & Liquid Phase Error (\%) \ \ \\
\hline
$\rho$         & 0.135158               & 0.094202                \\
$x_{CO2}$      & 0.064404               & 0.057355                \\
$x_{C1}$       & 0.089375               & 0.005375                \\
$x_{C2}$       & 0.223043               & 0.174558                \\
$x_{C3}$       & 0.364711               & 0.222567                \\
$x_{iC4}$      & 0.436977               & 0.235278                \\
$x_{C4}$       & 0.414238               & 0.191214                \\
$x_{iC5}$      & 0.414815               & 0.149893                \\
$x_{C5}$       & 0.392730               & 0.114561                \\
$x_{C6}$       & 0.314924               & 0.009726                \\
$x_{C7+}$      & 0.340858               & 0.511047                \\
\hline\hline
\end{tabular}
\end{table}

As can be seen by the low error values reported in Table \ref{tab10Cmpnt1DError}, the LBM correctly predicts the equilibrium for a ten-component mixture.

\begin{figure}[H]
    \centering
        \begin{subfigure}{0.49\textwidth}
            \centering
            \caption{}
            \includegraphics[width=\textwidth]{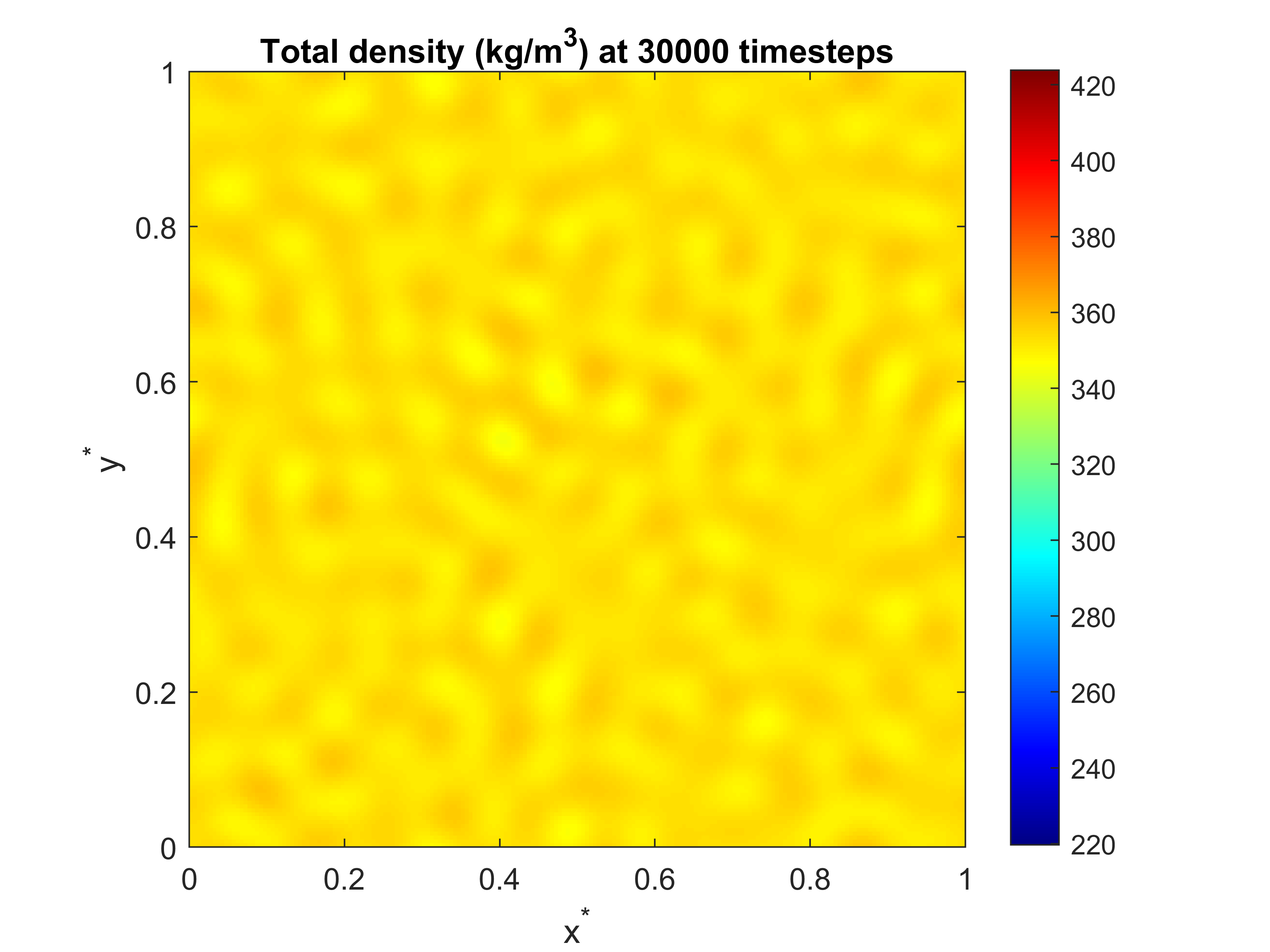}
        \end{subfigure}%
        \hfill
        \begin{subfigure}{0.49\textwidth}
            \centering
            \caption{}
            \includegraphics[width=\textwidth]{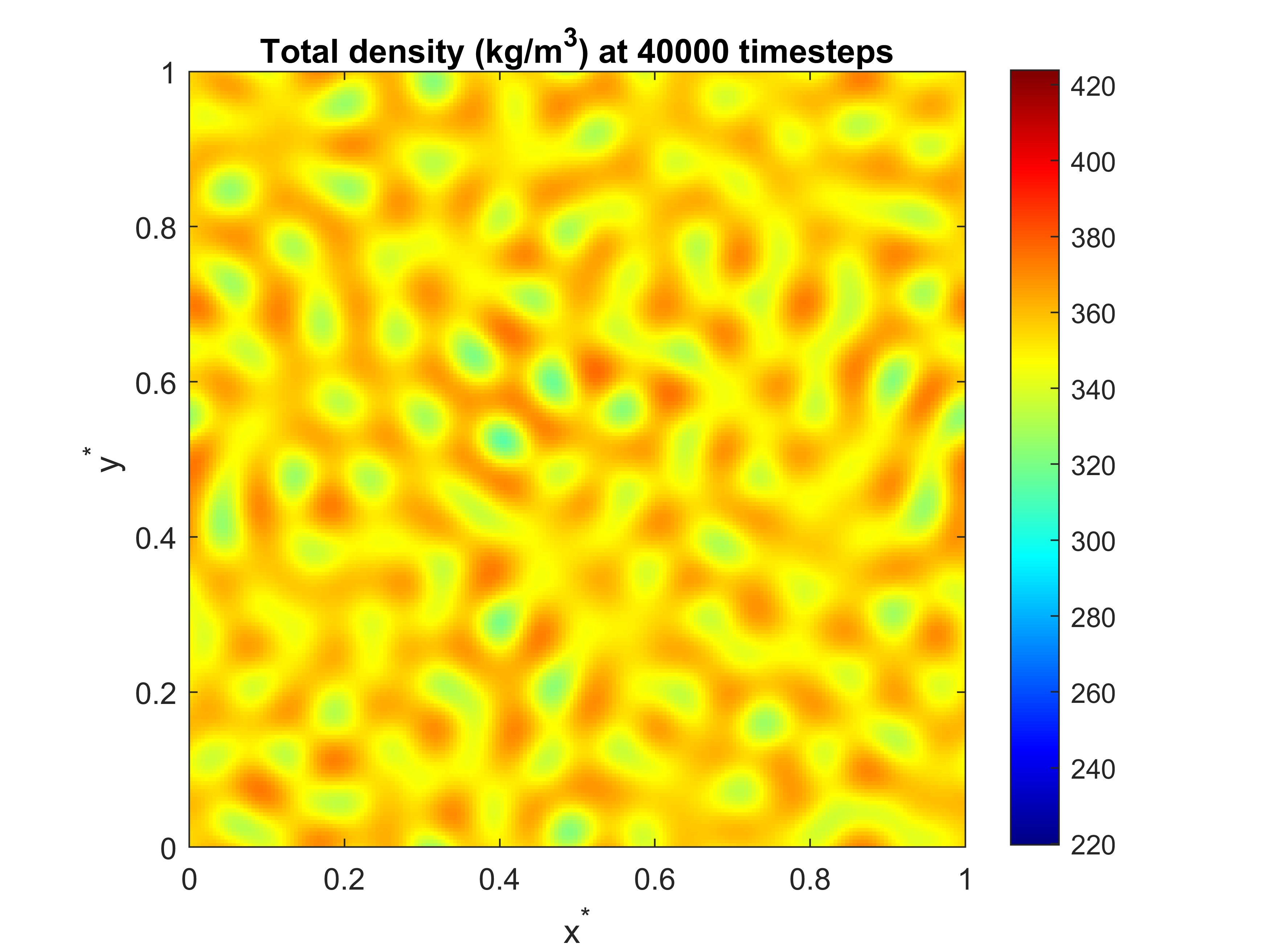}
        \end{subfigure}
        \hfill        
        \begin{subfigure}{0.49\textwidth}
            \centering
            \caption{}
            \includegraphics[width=\textwidth]{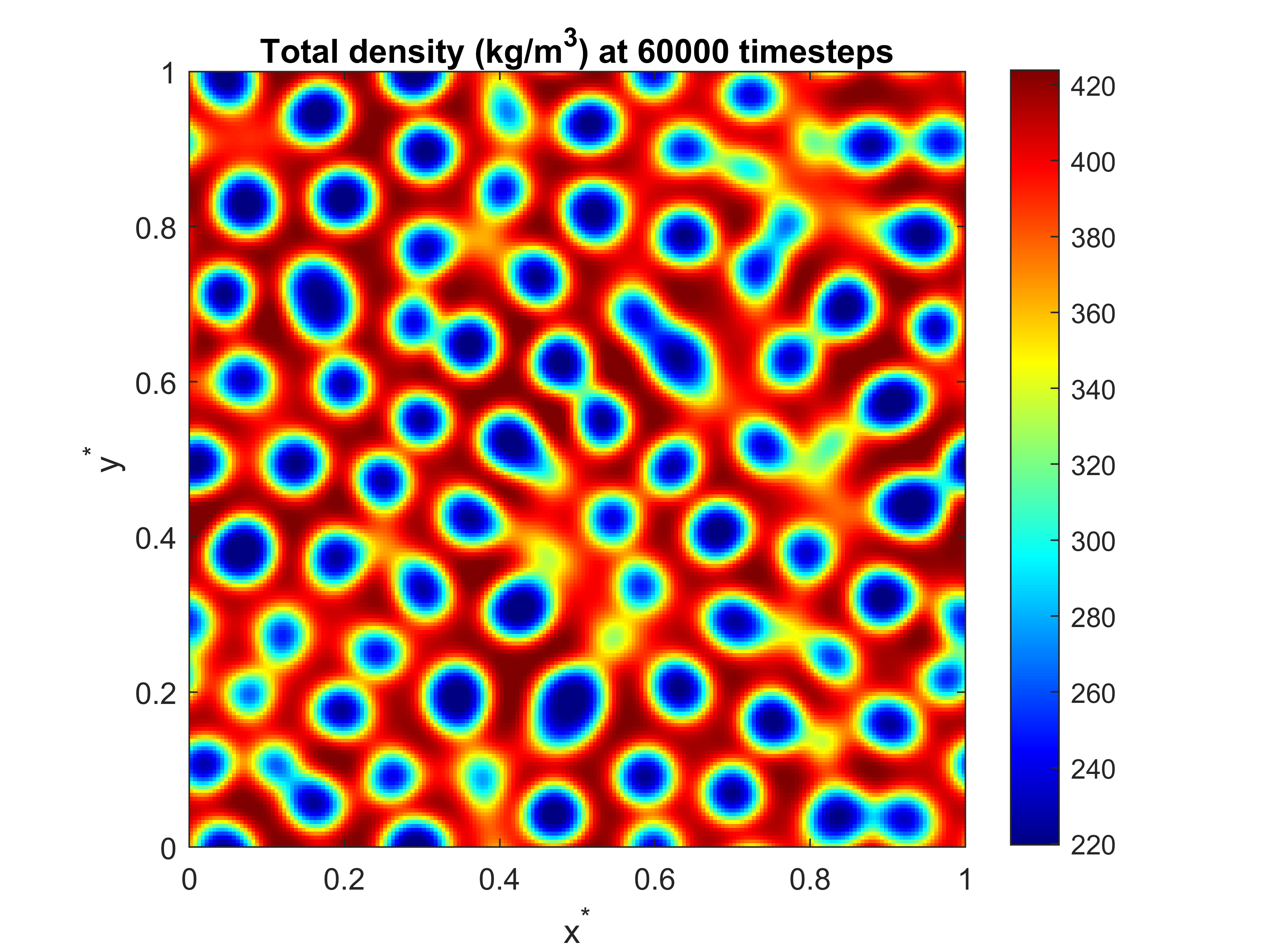}
        \end{subfigure}
        \hfill 
        \begin{subfigure}{0.49\textwidth}
            \centering
            \caption{}
            \includegraphics[width=\textwidth]{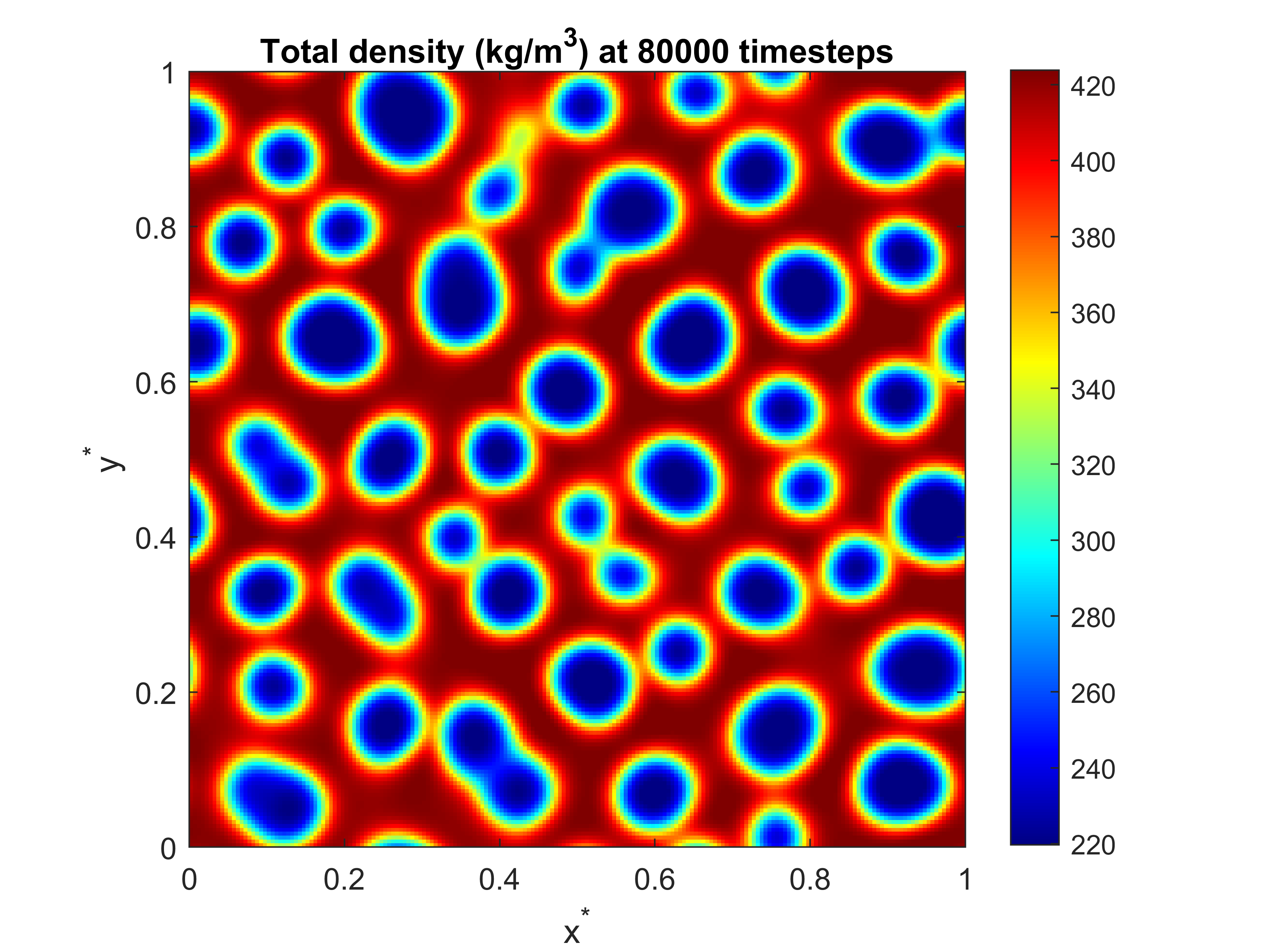}
        \end{subfigure}
        \hfill 
        \begin{subfigure}{0.49\textwidth}
            \centering
            \caption{}
            \includegraphics[width=\textwidth]{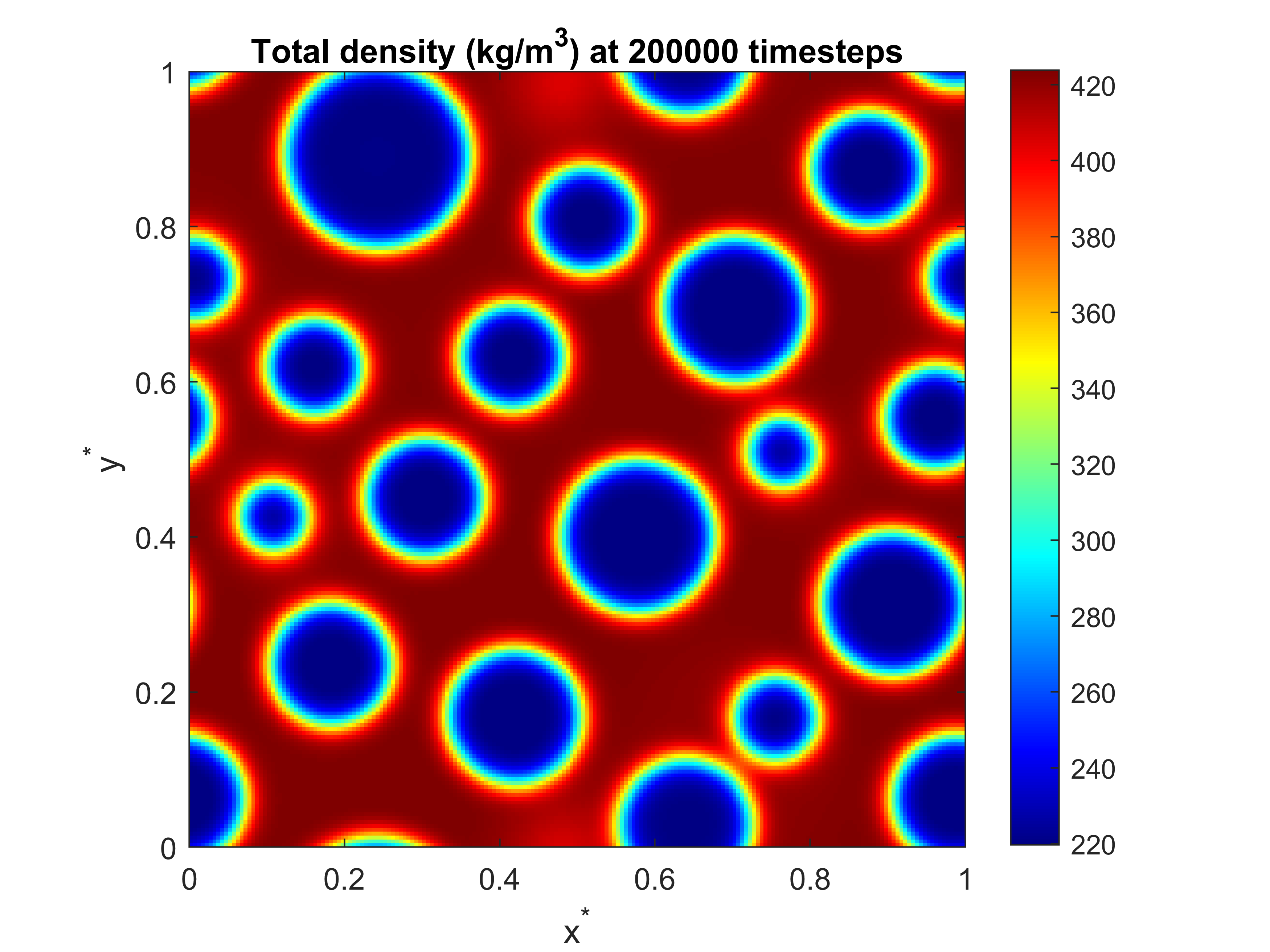}
        \end{subfigure}
        \hfill 
        \begin{subfigure}{0.49\textwidth}
            \centering
            \caption{}
            \label{fig10Cmpnt2DCase1f}
            \includegraphics[width=\textwidth]{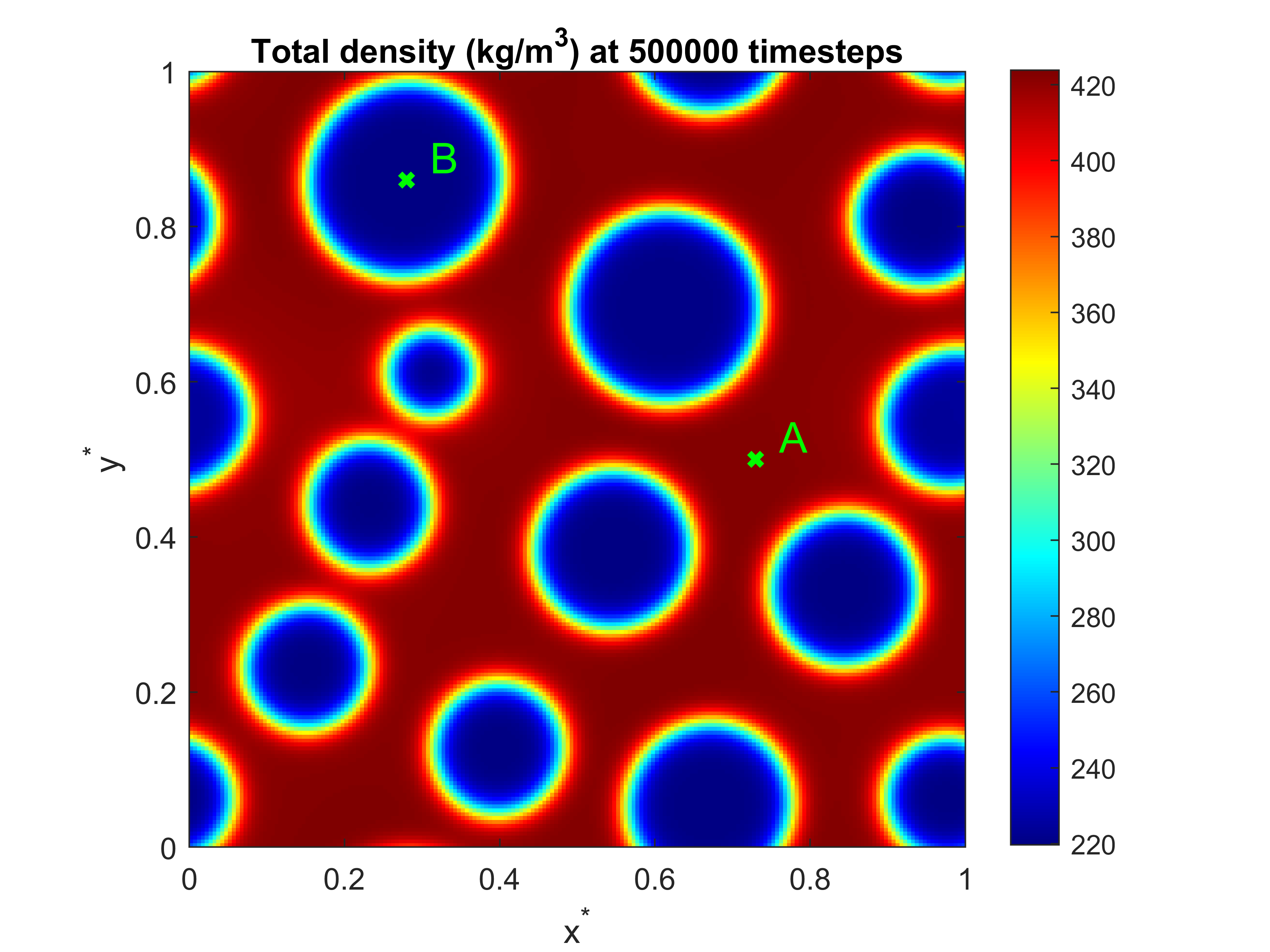}
        \end{subfigure}
        \hfill 

\caption{Spinodal decomposition of a ten-component system for Case 1 (liquid-dominated system). The figure shows the density profiles with dimensionless lengths: $x^*=x/n_x$ and $y^*=y/n_y$. These are at times (in lattice units): (a) 30,000, (b) 40,000, (c) 60,000, (d) 80,000, (e) 200,000, and (f) 500,000. The points marked A and B, in (f), represent the points in the liquid and vapor region, respectively, where component fugacities are measured.}
\label{fig10Cmpnt2DCase1}
\end{figure}

\begin{figure}[H]
    \centering
        \begin{subfigure}{0.49\textwidth}
            \centering
            \caption{}
            \includegraphics[width=\textwidth]{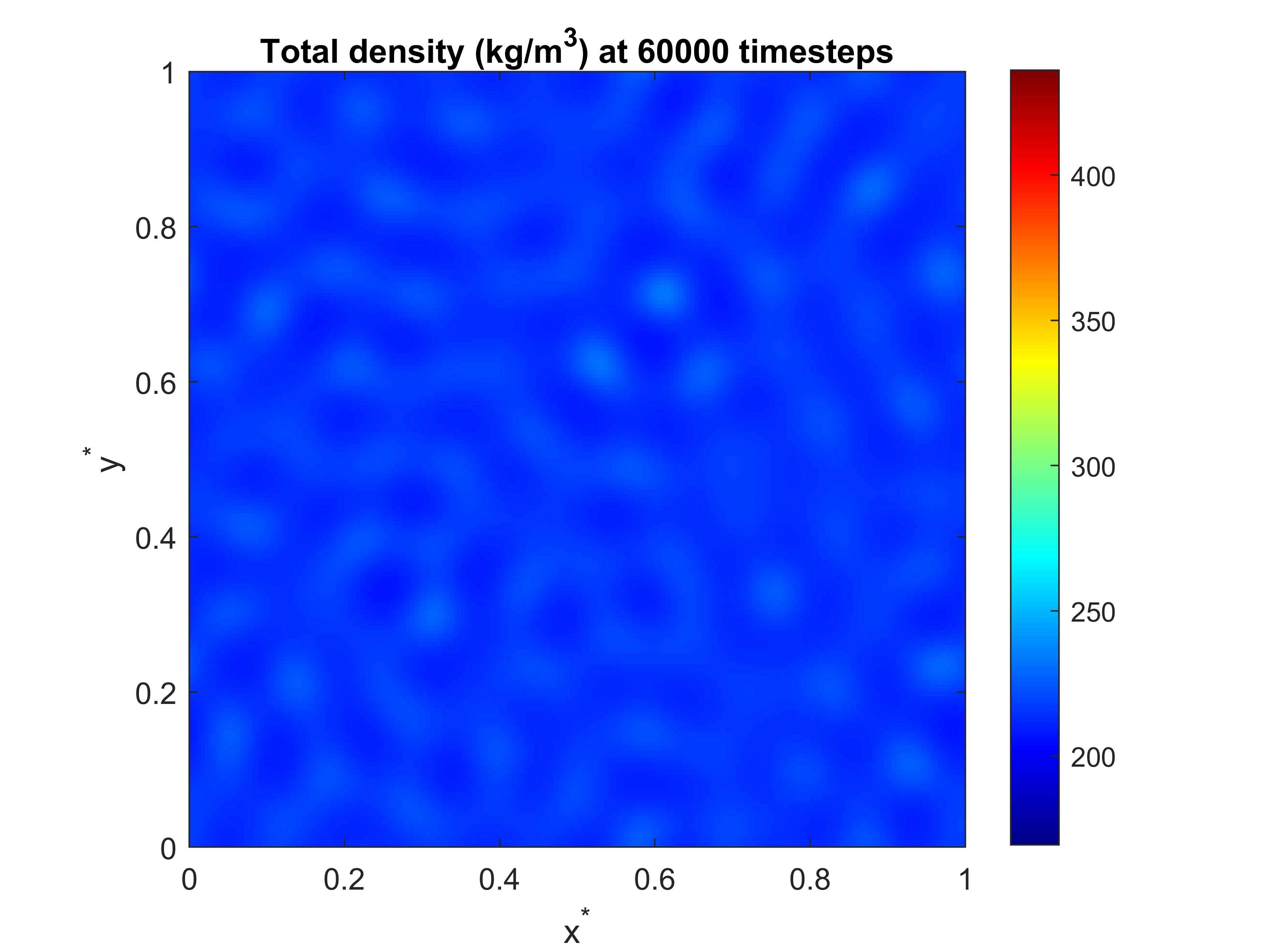}
        \end{subfigure}%
        \hfill
        \begin{subfigure}{0.49\textwidth}
            \centering
            \caption{}
            \includegraphics[width=\textwidth]{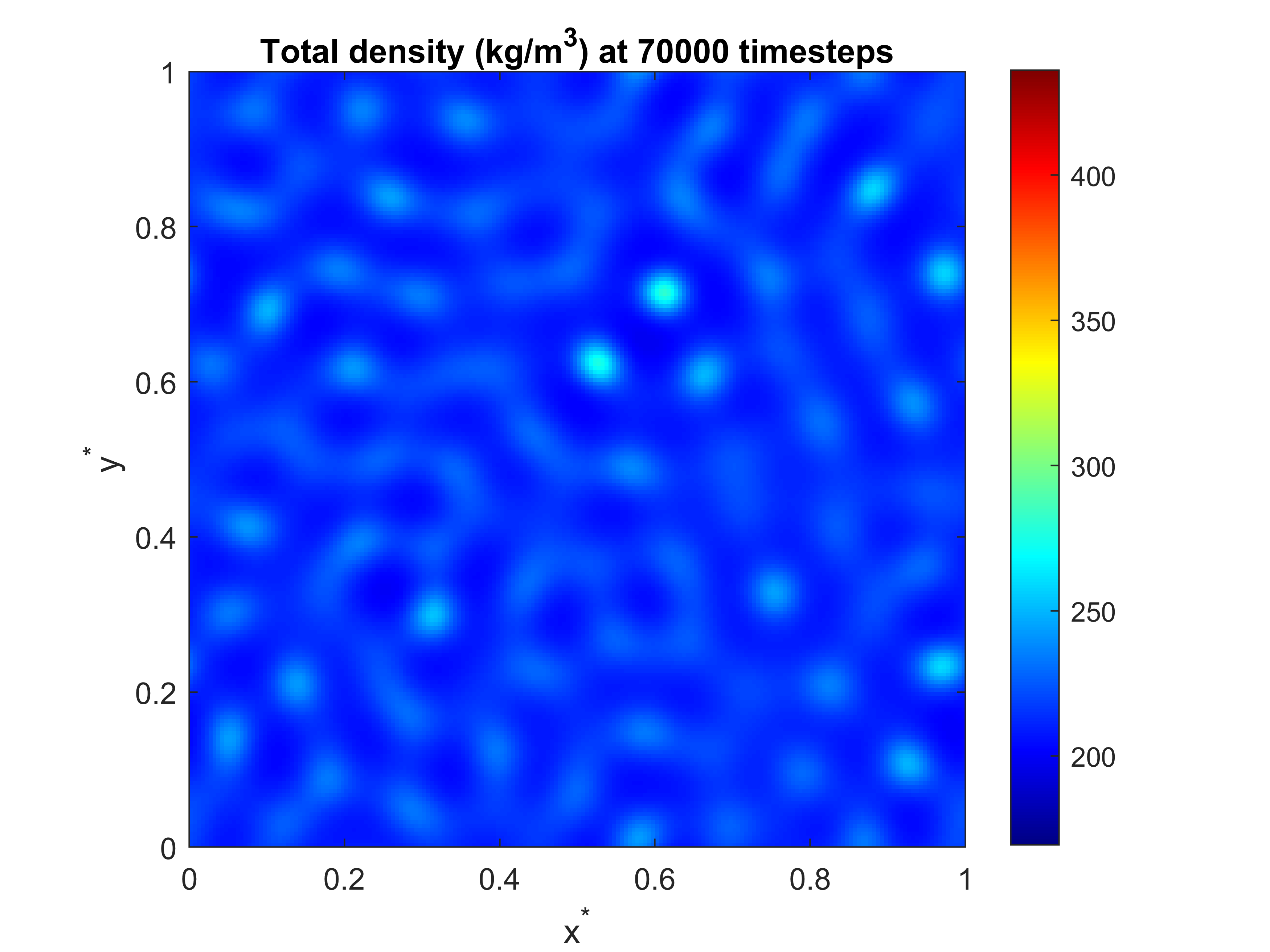}
        \end{subfigure}
        \hfill        
        \begin{subfigure}{0.49\textwidth}
            \centering
            \caption{}
            \includegraphics[width=\textwidth]{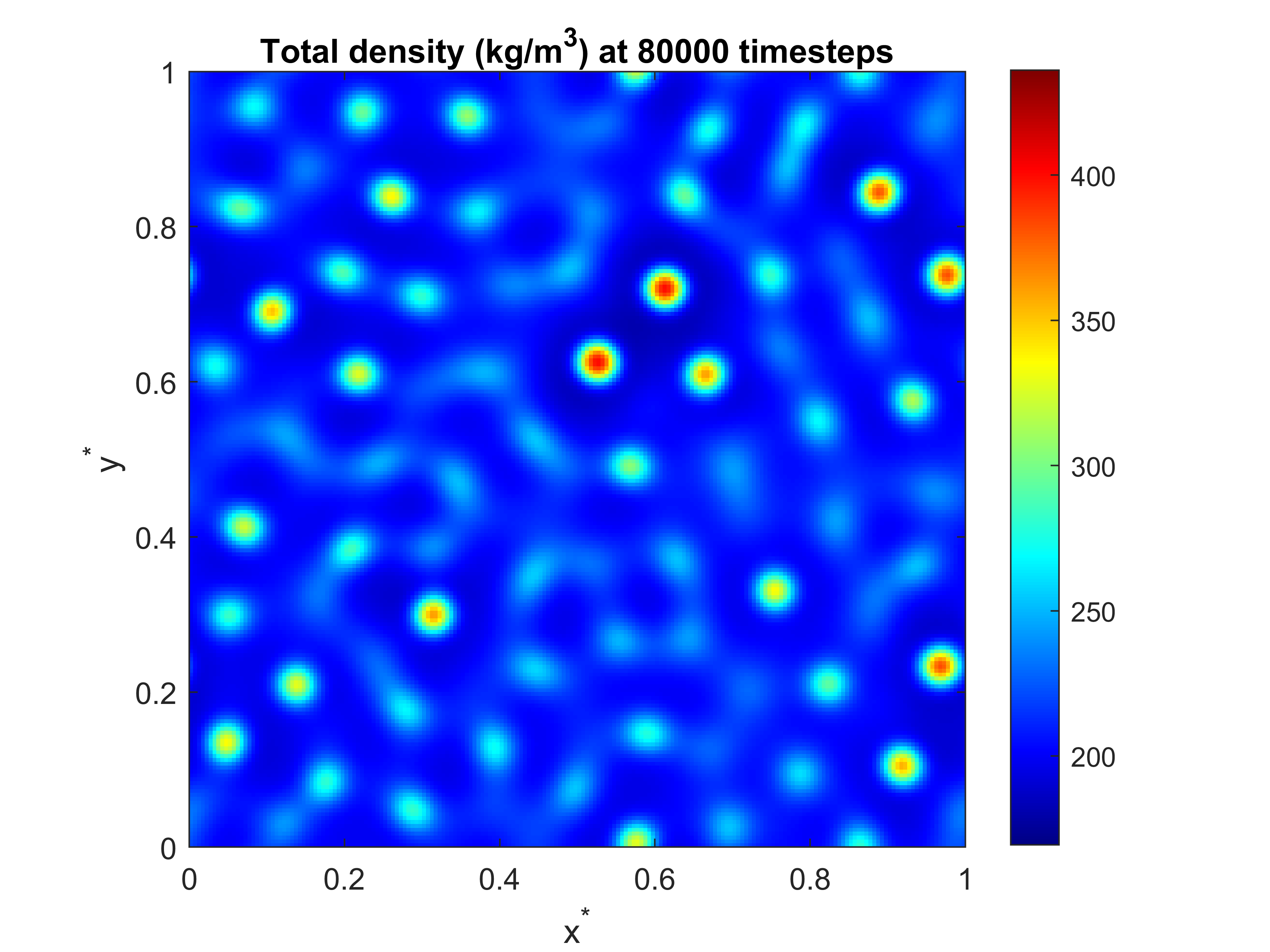}
        \end{subfigure}
        \hfill 
        \begin{subfigure}{0.49\textwidth}
            \centering
            \caption{}
            \includegraphics[width=\textwidth]{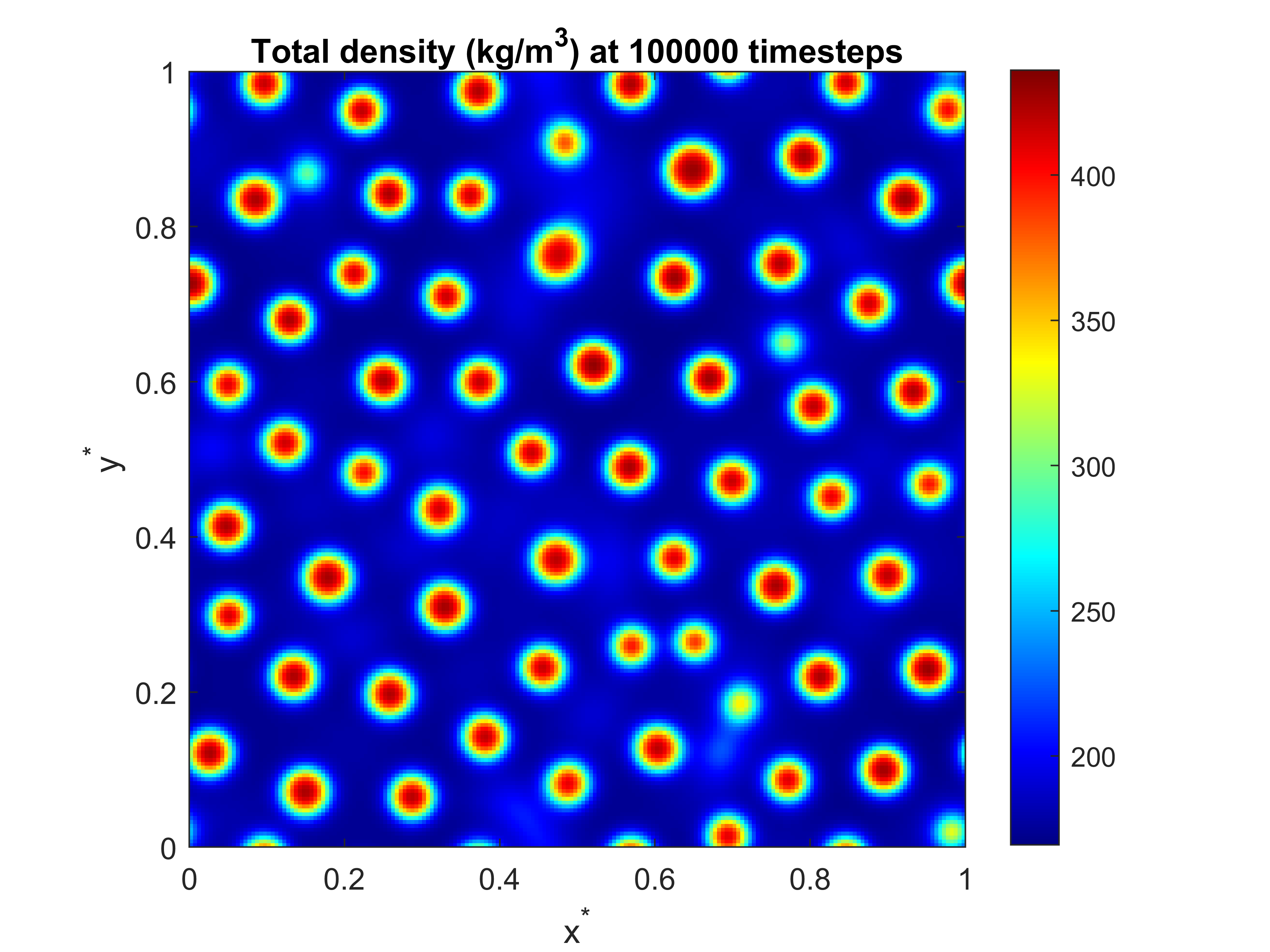}
        \end{subfigure}
        \hfill 
        \begin{subfigure}{0.49\textwidth}
            \centering
            \caption{}
            \includegraphics[width=\textwidth]{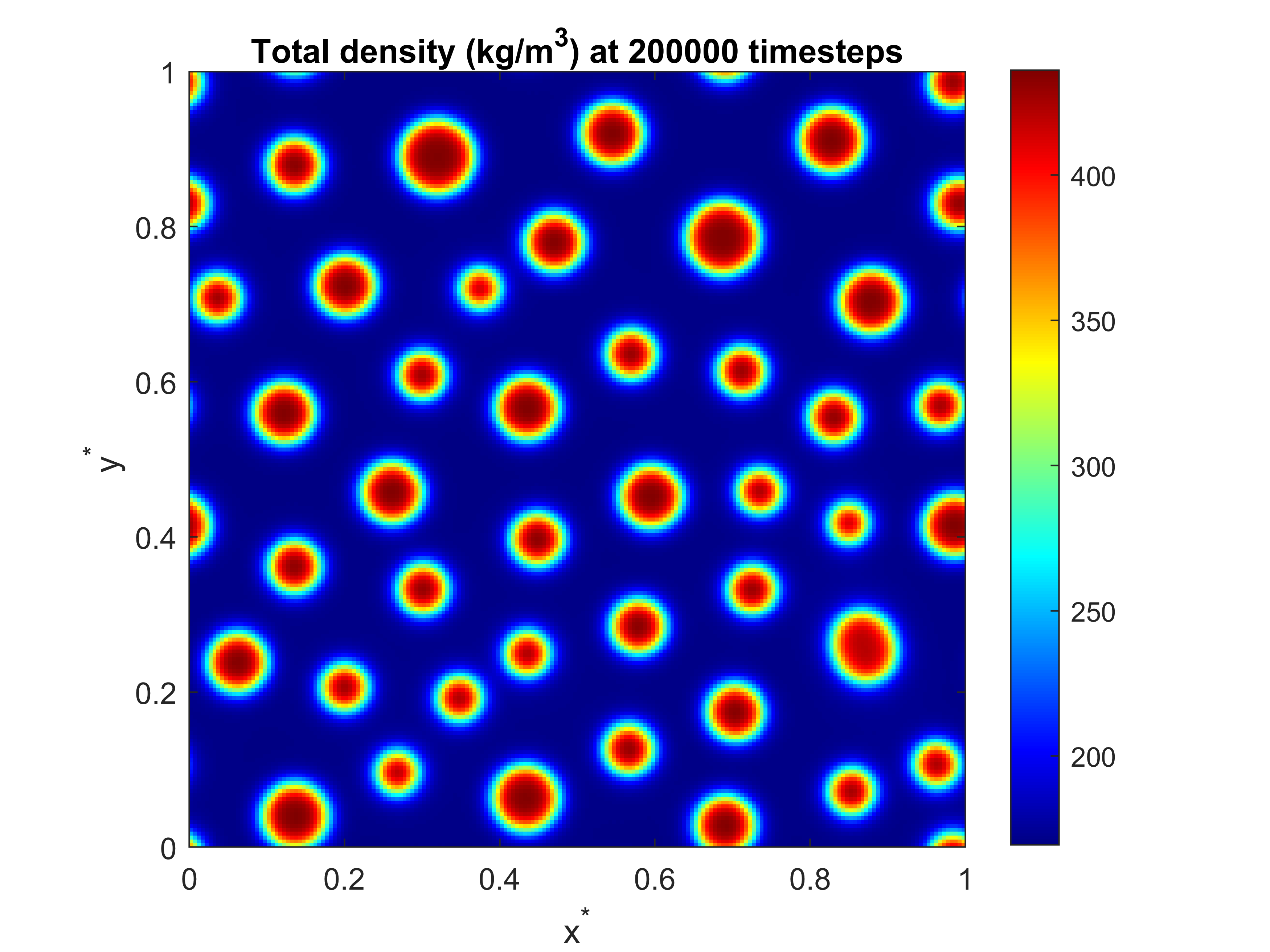}
        \end{subfigure}
        \hfill 
        \begin{subfigure}{0.49\textwidth}
            \centering
            \caption{}
            \label{fig10Cmpnt2DCase2f}
            \includegraphics[width=\textwidth]{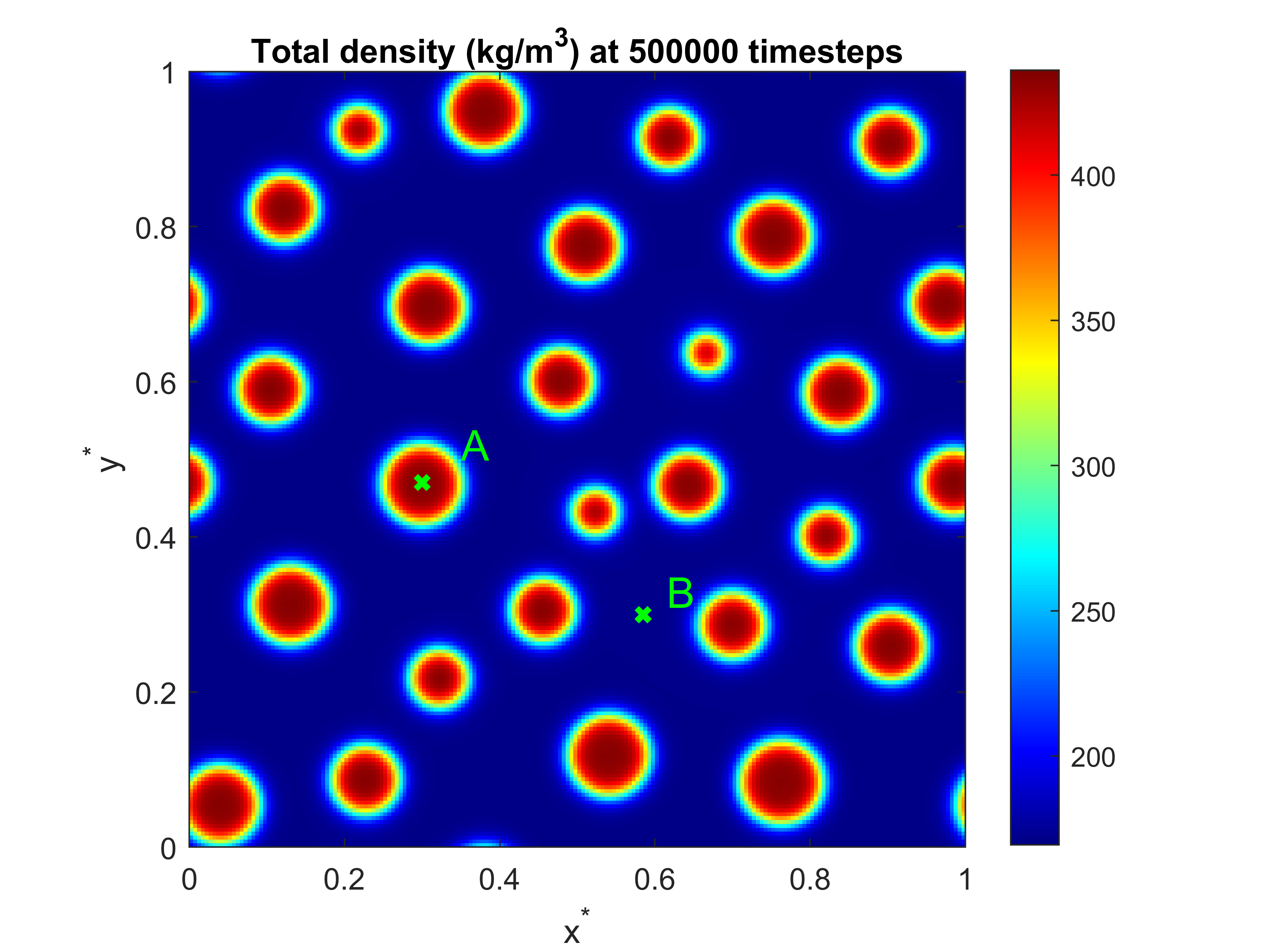}
        \end{subfigure}
        \hfill 

\caption{Spinodal decomposition of a ten-component system for Case 2 (vapor-dominated system). The figure shows the density profiles with dimensionless lengths: $x^*=x/n_x$ and $y^*=y/n_y$. These are at times (in lattice units): (a) 60,000, (b) 70,000, (c) 80,000, (d) 100,000, (e) 200,000, and (f) 500,000. The points marked A and B, in (f), represent the points in the liquid and vapor region, respectively, where component fugacities are measured.}
\label{fig10Cmpnt2DCase2}
\end{figure}

Next, we use the ten-component mixture to simulate the case of spinodal decomposition.
Thus far, the results in this paper have been shown for cases where the system is initialized close to equilibrium.
In this case, the mixture is initialized far from equilibrium, to test whether it still converges to the correct equilibrium values. The mixture initialized uniformly, and a small, random perturbation is introduced to the system. The test is conducted for two cases. In Case 1, a liquid-dominated system is formed with the liquid forming a continuous phase and the vapor forming bubbles. In Case 2, a vapor-dominated system is formed with the vapor forming a continuous phase and the liquid forming droplets. The different conditions are achieved by initializing Case 2 at a higher temperature than Case 1, with Case 1 at 325 K and Case 2 at 400 K.
The initial pressure for both cases is 150 bar, and the overall composition is provided in Table \ref{tabCompzi}.
Additionally, in both cases the relevant conversions between lattice units and physical units are established by fixing the universal gas constant and the attraction parameter, co-volume, and molar mass for C1 to the following values in lattice units: $R=1$, $a_{C1}=2/49$, $b_{C1}=2/21$, and $M_{C1}=1$. The relaxation time $\tau=1.0$, and the interfacial tension parameter for the reference component, chosen to be C1, is $\kappa_i^{ref}=0.02$. The size of the computational domain is $200\times 200$ ($n_x\times n_y$). Following a uniform initialization with a small, random perturbation, each case is run for 500,000 time-steps. The evolution of the system for Case 1 and Case 2 is illustrated in Figure \ref{fig10Cmpnt2DCase1} and \ref{fig10Cmpnt2DCase2}, respectively.

Figures \ref{fig10Cmpnt2DCase1} and \ref{fig10Cmpnt2DCase2} show the ten-component system spontaneously decomposing into two phases. Next, we verify the consistency of these results with thermodynamic predictions. In this section, we cannot rely on comparing the results to flash calculations as flash calculations assume a flat interface. Instead, we evaluate the consistency of our results by testing whether they adhere to the iso-fugacity criterion, which is valid for both flat and curved interfaces and serves as the basis of the flash calculation.
We recorded the fugacity of each component in the liquid and vapor phases for each case.  The fugacity measurements for the liquid phase were obtained at point A, and for the vapor phase, they were obtained at point B, marked on Figure \ref{fig10Cmpnt2DCase1f} for Case 1 and Figure \ref{fig10Cmpnt2DCase2f} for Case 2. The fugacity of each component in the vapor phase ($f_i^V$) and the liquid phase ($f_i^L$) are shown in Table \ref{tab10comp2Dcase1} for Case 1 and Table \ref{tab10comp2Dcase2} for Case 2. The ratio of the fugacity of each component in the liquid phase to its fugacity in the vapor phase is also summarized in the respective table. In Tables \ref{tab10comp2Dcase1} and \ref{tab10comp2Dcase2}, the fugacity ratios are very close to unity, indicating excellent agreement between the LBM results and thermodynamic predictions. Again, maximum deviations are associated with calculations for chemical components whose compositions are the smallest relative to others, for which the influence of computational round-off errors are bound to be more significant.

\begin{table}[H]
\caption{The fugacity of each component in the liquid phase, the fugacity of each component in the vapor phase, and their ratio, for Case 1.}
\label{tab10comp2Dcase1}
\centering
\begin{tabular}{lccc}
\hline \hline
Component & $f_i^L$ (bar) & $f_i^V$ (bar) & $f_i^L/f_i^V$ \\
\hline
CO2                           & 0.337479      & 0.337240      & 1.000707      \\
C1                            & 108.194561    & 108.180227    & 1.000133      \\
C2                            & 8.455564      & 8.450527      & 1.000596      \\
C3                            & 2.201804      & 2.199853      & 1.000887      \\
iC4                           & 0.138429      & 0.138273      & 1.001131      \\
C4                            & 0.393377      & 0.392999      & 1.000961      \\
iC5                           & 0.051556      & 0.051503      & 1.001025      \\
C5                            & 0.076134      & 0.076066      & 1.000895      \\
C6                            & 0.041749      & 0.041727      & 1.000523      \\
C7+                           & 0.007155      & 0.007176      & 0.997187      \\
\hline \hline
\end{tabular}
\end{table}

\begin{table}[H]
\caption{The fugacity of each component in the liquid phase, the fugacity of each component in the vapor phase, and their ratio, for Case 2.}
\label{tab10comp2Dcase2}
\centering
\begin{tabular}{lccc}
\hline \hline
Component & $f_i^L$ (bar) & $f_i^V$ (bar) & $f_i^L/f_i^V$ \\
\hline
CO2       & 0.434208      & 0.432687      & 1.003514      \\
C1        & 104.847165    & 104.816434    & 1.000293      \\
C2        & 13.950388     & 13.959195     & 0.999369      \\
C3        & 5.419398      & 5.426715      & 0.998652      \\
iC4       & 0.454399      & 0.455480      & 0.997627      \\
C4        & 1.438833      & 1.441916      & 0.997862      \\
iC5       & 0.252597      & 0.253289      & 0.997270      \\
C5        & 0.408427      & 0.409779      & 0.996701      \\
C6        & 0.323627      & 0.325112      & 0.995431      \\
C7+       & 0.221562      & 0.224617      & 0.986397      \\
\hline \hline
\end{tabular}
\end{table}

\section{Discussion and Conclusion}
\label{secConclusion}

This paper presents a thorough analysis of partially-miscible mixtures with multiple components. Previous studies using the LBM are limited in the number of components used, particularly for partially-miscible systems. In this work, we employ the recently published fugacity-based LBM and introduce a minor approximation to remove the component restrictions on LBM models. Our model provides results in excellent agreement with thermodynamic predictions, regardless of the number of components used.

We begin by demonstrating the compliance of our model with the Young-Laplace equation. This is achieved by simulating a droplet composed of three components and showing the proportionality of the capillary pressure with the inverse droplet radius at various conditions. Next, we present a flat interface VLE case for mixtures with a range of component numbers from one to six. Our results agree with thermodynamic predictions. We also demonstrate that the computational time of LBM simulations scales linearly with the number of components. Further, we perform a deeper analysis of the phase behavior of ternary systems, exploring a wide range of temperature, pressure, and overall composition conditions to produce various characteristic ternary diagrams. Our model is also demonstrated to be unrestricted in the number of phases, as we simulate a three-component three-phase equilibrium case. Finally, we conclude our paper by presenting simulations of a ten-component hydrocarbon mixture obtained from literature. We perform flat interface VLE and spinodal decomposition cases for this mixture and demonstrate excellent agreement with thermodynamics.

This paper has highlighted, examined, and demonstrated a key contribution to state-of-the-art LB simulation: multiphase LB can be straightforwardly and confidently extended to any number of chemical components while maintaining full agreement with multicomponent, multiphase thermodynamic principles using the proposed approach. To the best of the authors' knowledge, the largest number of components ever utilized in LBM simulations for a partially-miscible system has been three \cite{Ridl2018,Gong2014,Bao2013} without strict compliance with multicomponent thermodynamics. Maximum number of chemical components has gone beyond three, and reached up to five components, for immiscible systems only \cite{Zheng2020MC}. In contrast, our simulations include up to ten-component partially-miscible systems, all of which achieve precise thermodynamic consistency without any corrections or tuning. Even for three-component partially-miscible systems, previous studies in the literature have been limited in scope in terms of the thermodynamic domain being explored. Previous studies explored a single condition/case, restricted cases to vdW fluids only (vdW cubic EOS is known for lack of quantitative agreement with experimental data), and/or presented approaches which lacked consistency with thermodynamics \cite{Ridl2018,Gong2014,Bao2013}. In this work, we have presented a more comprehensive study of ternary systems over a wide range of temperature, pressure, and overall composition using the PR EOS, known for its reliability for hydrocarbon systems. Our simulations generate several characteristic ternary diagrams for this system, all of which are fully consistent with thermodynamic predictions. This expansion of the scope of LBM simulations in terms of the number of components and the range of conditions studied represents a significant step towards a more comprehensive modeling and understanding of complex mixture flow behavior.

\begin{acknowledgments}
Funding support from the William A. Fustos Family Professorship in Energy and Mineral Engineering at the Pennsylvania State University is gratefully acknowledged.
\end{acknowledgments}

\bibliography{References}

\end{document}